\pdfoutput=1
\documentclass[
reprint,
 amsmath,amssymb,
 aps,
]{revtex4-1}

\usepackage{graphicx}
\usepackage{dcolumn}
\usepackage{bm}
\usepackage{bbold}

\begin{document}

\preprint{APS/123-QED}
\title{Efficient evaluation of high-order moments and cumulants in tensor network states}

\author{Colin G. West}

\affiliation{%
C. N. Yang Institute for Theoretical Physics and Department of Physics and Astronomy, Stony Brook University, Stony Brook, NY 11794}%


\author{Artur Garcia-Saez}
\affiliation{%
C. N. Yang Institute for Theoretical Physics and Department of Physics and Astronomy, Stony Brook University, Stony Brook, NY 11794}%

\author{Tzu-Chieh Wei}
\affiliation{%
C. N. Yang Institute for Theoretical Physics and Department of Physics and Astronomy, Stony Brook University, Stony Brook, NY 11794}%


\date{\today}

\begin{abstract}
We present a numerical scheme for efficiently extracting the higher-order moments and cumulants of various operators on spin systems represented as tensor product states, for both finite and infinite systems, and present several applications for such quantities. For example, the second cumulant of the energy of a state, $\langle \Delta H^2 \rangle$, gives a straightforward method to check the convergence of numerical ground-state approximation algorithms. Additionally, we discuss the use of moments and cumulants in the study of phase transitions. Of particular interest is the application of our method to calculate the so-called BinderÕs cumulant, which we use to detect critical points and study the critical exponent of the correlation length with only small finite numerical calculations. We apply these methods to study the behavior of a family of one-dimensional models (the transverse Ising model, the spin-1 Ising model, and the spin-1 Ising model in a crystal field), as well as the two-dimensional Ising model on a square lattice. Our results show that in one dimension, cumulant-based methods can produce precise estimates of the critical points at a low computational cost, and show promise for two-dimensional systems as well. 
\end{abstract}

\maketitle


\section{\label{sec:level1}Introduction}


The understanding of quantum many-body systems is one of the foremost goals of modern quantum physics. In addition to various analytical approaches, a new set of numerical tools has emerged for this purpose in recent years, based on tensor network representations of such systems \cite{TN1, MPSReview, MPSReview2, OrusReview}. These techniques take advantage of the so-called ``area law" for entanglement entropy, obeyed by the low-energy eigenstates of gapped Hamiltonians with local interactions. Tensor networks naturally embody this entanglement structure, dramatically simplifying the degrees of freedom required to describe such states. 
Initially introduced in the context of gapped one-dimensional systems, where they are referred to as ``matrix product states" (MPS) \cite{OriginalMPS1, OriginalMPS2, OriginalMPS3}, tensor network methods rose to even greater prominence when it was realized that the celebrated ``Density Matrix Renormalization Group" (DMRG) technique \cite{DMRGWhite1, DMRGWhite2} could be reformulated in terms of an MPS \cite{DMRGMPS1, DMRGMPS2, DMRGMPS3, DMRGMPSReview}. Additional algorithms soon followed DMRG, such as `Time Evolving Block Decimation" (TEBD) \cite{TEBD}, as well as the infinite-system analogues ``iDMRG" and ``iTEBD" \cite{iDMRG, iTEBD}. Since then, tensor network methods have also been readily applied to critical systems be means of the ``Multiscale Entanglement Renormalization Anzatz" (MERA) \cite{MERAOriginal, MERA2}, as well as to higher-dimensional systems, where they are generally termed ``tensor network states," or ``projected-entangled pair states" (PEPS) \cite{PEPSOriginal, MurgMPO, iPEPS,PEPS1, PEPS2, TreeTN1, TreeTN2}. A principle goal of these algorithms is to obtain precise approximations to ground-state wave functions, which can be used for many purposes. For example, one can compute various quantities and observables in an effort to detect phase transitions, a central problem in many-body physics \cite{PT1, PT2, PT3, PT4, PT5}. 

In the Landau symmetry-breaking paradigm for phase transitions, one first looks for an operator $M$ whose expectation value $\langle M \rangle$ can serve as the order parameter, i.e. a quantity whose behavior changes sharply across a critical point. When this order parameter is represented by a local operator, it can be computed efficiently on a tensor network state \cite{MPSReview}. But while an expectation value is the most straightforward piece of information associated with an operator and a state, there is considerably more information available which one may want to compute. For instance, one may wish to study the higher moments of the operator,  $\mu_n =  \langle M^n \rangle $. A related set of quantities called ``cumulants," typically labelled $\kappa_n$, is also frequently of interest. An obvious example is the variance of the operator $\langle \Delta M^2 \rangle$, which is simply the second cumulant $\kappa_2 = \mu_2 - \mu_1^2$. Even more important to the search for phase transitions is the so-called ``Binder cumulant," first introduced by Kurt Binder in 1981 in a study of the classical Ising Model \cite{Binder1981}. In many settings, such as thermal or disordered systems, it is considered to be one of the most accurate and reliable means of detecting a critical point \cite{BC1, BC2, BC3}, and it has since been applied to a wide variety of models \cite{BinderMonteCarlo, Selke1, Selke2, Selke3, BinderApp1, BinderApp2, BinderApp3, BinderApp4}.  

Computing these higher order moments and cumulants, however, is less straightforward. Direct calculation quickly becomes impractical for large $n$, since the number of terms to evaluate can be exponential in $n$. In a classical system with a Hamiltonian $H_0$, one might define $H(\lambda) = H_0 + \lambda M$, and relate the higher moments of $M$ to the derivatives of an associated partition function, using $\langle M^n \rangle = (\beta \frac {\partial}{ \partial \lambda})^n Tr(e^{-\beta H(\lambda)})$. In quantum systems, however, this equation only holds when $[H_0, M] = 0$, which is not true for a wide variety of physically interesting cases. Because of these barriers to direct calculation, usage of techniques such as the powerful Binder cumulant has in the past been generally confined to studies based on quantum Monte Carlo \cite{BinderMonteCarlo}. 

The question naturally arises whether these quantities can be efficiently and systematically evaluated using the elegant structure of a tensor network state. Here, we demonstrate that the answer is yes. The feasibility of using matrix product states for computing the second moment of Hamiltonians was already pointed out by I. McCulloch in the context of DMRG \cite{IanMPO}, via the technique of so-called ``matrix product operators" (MPO) \cite{MurgMPO}. In this work, we propose a simple and efficient method to allow all general moments and cumulants to be evaluated for tensor network states, based on moment-generating and cumulant-generating functions. We demonstrate the calculation of moments and cumulants for finite one-dimensional states, and show that the method can also be used for per-site cumulants in the case of an infinite system. We also show how the techniques naturally generalize to finite systems in higher dimensions. These methods have a variety of useful applications which are demonstrated at length, including the use of the Binder and other cumulants to detect critical points to relatively high precision at a low numerical cost. We also apply the second cumulant of the energy to examine the convergence of numerical methods based on imaginary time evolution.

This paper is organized as follows: In Section II we review moments and cumulants, in particular presenting the Binder cumulant and some of its applications. In Section III, we very briefly review the MPS formalism and discuss how to compute certain simple expectation values. Section IV demonstrates how to use these expectation values to efficiently compute the moments and cumulants of general operators on an MPS. Section V contains examples of the method as applied to three different spin-chain models (the transverse Ising model, the spin-1 Ising model, and the spin-1 Ising model in a crystal field), as well as a demonstration of the method as applied to a two dimensional system (the transverse Ising model on a square lattice). Our results are summarized in Section VI.

\section{Moments, Cumulants, and The Binder Cumulant}

A state $|\psi \rangle$ and an operator $M$ collectively imply a probability distribution: the probability density function of $\psi$ in $M$-space. The expectation value $\langle M \rangle$ specifies the central value of the distribution, while the complete set of ``Moments" defines the entire the shape \cite{Moments}. The $n^{th}$ moment of the distribution is defined to be  $\mu_n =  \langle M^n \rangle $; the first moment $\mu_1$ is the expectation value $\langle M \rangle$ itself.

The cumulants of the distribution, $\kappa_n$, form an alternative but equivalent way of specifying its shape. These cumulants contain, in total, the same information as the moments; a complete set of either moments or cumulants completely specifies the distribution. Indeed, the $n^{th}$ cumulant can always be expressed as a polynomial combination of the first $n$ moments, and vice versa \cite{MomentsToCumulants}. For example, as we have noted above, the second cumulant of the distribution, is the distribution's variance, defined by

\begin{equation}
\kappa_2 = \mu_2 - \mu_1^2.
\end{equation}

The third cumulant $\kappa_3$ gives the distribution's skewness, and is related to the first three moments by

\begin{equation}
\kappa_3 = \mu_3 - 3\mu_2\mu_1 + 2\mu_1^3.
\end{equation}

Similarly, the fourth cumulant $\kappa_4$ is related to the kurtosis, and is given by

\begin{equation}
\kappa_4 = \mu_4 - 4\mu_3\mu_1 - 3\mu_2^2+12\mu_2\mu_1^2-6\mu_1^4.
\end{equation}

Although moments and cumulants are properly defined with respect to a distribution and hence depend on both $M$ and $| \psi \rangle$, when $| \psi \rangle$ is general or clear from context we shall refer to $\mu_n$ ($\kappa_n$) as the ``$n^{th}$ moment (cumulant) of $M$".

\subsection{Binder's Cumulant}

The aforementioned Binder Cumulant is a particularly useful quantity in the study of critical points and phase transitions. For some system with some known order parameter $M$, for example a total magnetization $\sum_j \sigma_j$ or a staggered magnetization $\sum_j (-1)^j \sigma_j$, Binder's cumulant represents a modified version of that parameter's 4th cumulant. Though some slight variations exist in the definition, generally, it is given by

\begin{equation}\label{eq:BinderDef}
U_4 = 1- \frac{ \langle M^4 \rangle}{3 \langle M^2 \rangle^2}.
\end{equation} 

The utility of the Binder cumulant arises from the special features of its length dependance. The behavior of the Binder cumulant at a critical point depends only weakly on the size of the system, and elsewhere, its behavior with respect to the system size differs depending upon the phase. For example, below the critical point in a symmetry-breaking magnetic phase the cumulant will increase with the length of the system, but above the critical point, with symmetry unbroken, it decreases instead. The result is that, when curves of the Binder Cumulant vs temperature are plotted for various lengths, the critical point is indicated by a simultaneous crossing. Typically, because the behavior at the critical point is already approximately universal, only a set of relatively small system sizes need be considered, eliminating the need for complicated extrapolations of very large systems to the thermodynamic limit. 

The Binder cumulant also gives access to the critical exponent of the correlation length, by means of traditional finite size scaling techniques in which one seeks to ``collapse" the data. \cite{Binder1981, BinderFSS}. Up to some small finite size corrections (which become increasingly suppressed as the system size increases), the cumulants show a standard functional form, \cite{BinderFSS}

\begin{equation}
U_4(L, B) = \tilde U \left(L^{1/\nu} (B-B_c) \right),
\end{equation}
where $\nu$ is the usual critical exponent. A plot of $U_4$ vs $L^{1/\nu} (B-B_c)$ should therefore appear essentially independent of $L$, since all of the length-dependance has been absorbed into the independent variable of the plot. Hence,  $B_c$ and $\nu$ can be treated as free parameters, and varied until this length-independence is optimized; for example, one could seek to minimize the total absolute square distance between the curves for a variety of lengths $L$.

\section{Matrix Product States and Expectation Values}

\subsection{Finite-length chains}
To demonstrate how to efficiently compute quantities such as the Binder cumulant for a system represented by an MPS, we must first review the nature of the MPS representation itself. Consider first a simple 1-dimensional spin chain of length $L$ with periodic boundary conditions. As an MPS, this state will be expressed as

\begin{equation}\label{eq:MPS}
| \psi \rangle = \sum_{s} Tr( A_1^{(s_1)}A_2^{(s_2)}...A_L^{(s_L)})| s_1 s_2 ... s_L \rangle.
\end{equation}

In other words, a rank-three tensor ``$A_j$" has been associated with each site ``$j$". One index of this tensor, denoted $s_j$ above, remains free and represents the physical degrees of freedom at site $j$. The other two indices, typically termed the ``virtual indices", are contracted with the virtual indices of the neighboring tensors $A_{j-1}$ and $A_{j+1}$. The dimension of these virtual indices, often labelled $\chi$, is called the ``bond dimension" of the MPS. The choice of $\chi$ represents a numerical parameter which can be adjusted to suit the requirements of the context: smaller values are of course less numerically expensive, but larger values can allow the MPS to more accurately represent the features of the state, particularly in systems with long correlation lengths.

In the one-dimensional case, since A is rank-3, for any fixed value of $s_j$ the two remaining virtual indices simply represent a matrix. For this reason, the labels of the virtual indices are often suppressed, as in Eq. (\ref{eq:MPS}), with the contraction represented by the expression $A_{j-1}^{(s_{j-1})}A_{j}^{(s_{j})}A_{j+1}^{(s_{j+1})}$ in the same manner that one would write an ordinary product of matrices.

Let us now consider how to calculate the expectation value of an operator with respect to a matrix product state. Consider first the case of a simple operator which is given by a tensor product of on-site operations. For such an operator, of the form 

\begin{equation}\label{eq:ProdOp}
Q = \bigotimes_j Q_j, 
\end{equation}
the expectation value $\langle Q \rangle$ is given by

\begin{equation}
\langle \psi | Q | \psi \rangle = \sum_{s_j, s'_j} \left[ Tr \left( \prod _{j= 1}^L  A_j^{(s_j)} \otimes A_j^{*(s'_j)}\right) \prod_{j = 1}^L \langle s'_j | Q_j | s_j \rangle \right],
\end{equation} 
or equivalently, 

\begin{equation}
 =  Tr \left[ \prod_{j = 1} ^L \sum_{s_j, s'_j} \left(  A_j^{(s_j)} \otimes A_j^{*(s'_j)} \right)  \langle s'_j | Q_j | s_j \rangle \right].
\end{equation} 

From this expression, it is clear that, up to normalization, the expectation value is simply a trace over a set of $L$ ``transfer matrices"; i.e.

\begin{equation}\label{eq:BinderExpt}
\langle Q \rangle = \frac{1}{\langle \psi | \psi \rangle} Tr \left( \prod_{j = 1}^{L}T_j \right),
\end{equation}
where the $T$'s are defined as 

\begin{equation}\label{eq:Transfer}
T_j \equiv \sum_{s_j, s'_j} \left(  A_j^{(s_j)} \otimes A_j^{*(s'_j)} \right)  \langle s'_j | Q_j | s_j \rangle.
\end{equation}

\begin{figure}[ht]
\includegraphics[width = 90mm]{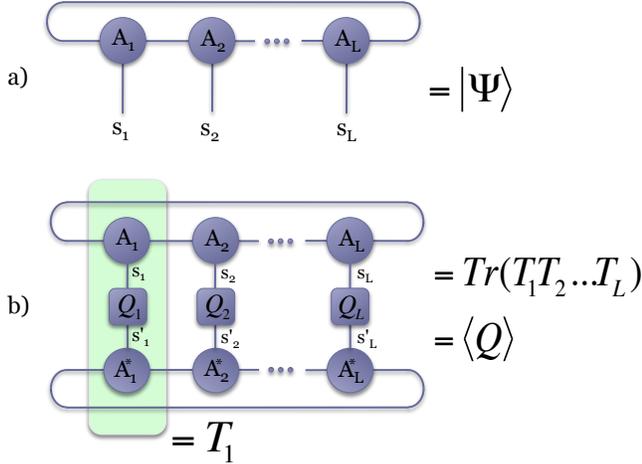}
\caption{\label{fig:Expectation} Graphical notation demonstrating the structures of matrix product states. In this notation, a shape represents a tensor, and a line represents an index. Connected lines between shapes represent contracted indices between tensors. (a) A finite spin chain state $| \psi \rangle$ represented as a matrix product state. The state is specified by the set of rank-three tensors $\{ A_j \}$, with the physical degrees of freedom $s_j$ left open. (b) The expectation value of a product operator $Q = \otimes_j Q_j$ with respect to $| \psi \rangle$. Each $Q_j$ acts locally on only one site. The total expectation value can be thought of as a trace over a product of transfer matrices $T_j$, defined in Eq. (\ref{eq:Transfer}). An example of an individual transfer matrix, $T_1$ is highlighted.}
\end{figure}

This procedure is also demonstrated in graphical notation in Fig.~\ref{fig:Expectation}. The norm of the state can be fixed in a similar fashion, by evaluating a transfer matrix for the special case where $Q_j = \mathbb{1}$. 

Tensor products of few-body operators can be handled in a similar fashion by grouping the relevant sites. More general operators are simply evaluated by decomposing them into a sum of tensor products. Considerably more detail on the general process of taking expectation values can be found in the now-extensive body of literature on matrix product states, \cite{OrusReview, MPSReview, CanonicalForm}. For our purposes, however, it will be sufficient to be able to evaluate operators of the simple form in Eq. (\ref{eq:ProdOp}).

\subsection{Infinite-length chains}

One significant advantage of tensor network algorithms is how easily they allow one to directly study certain infinite systems. This can be done for systems with some form of translation invariance, which can be completely represented by their unit cells. For example, consider an one-dimensional infinite system possessing translation invariance with respect to a unit cell of length $\ell$. Represented as a matrix product, the state is of the form given by Eq. (\ref{eq:MPS}), but with the further restriction that not all tensors $A_i$ are distinct, and instead repeat every $\ell$ sites. A state with only a one-site unit cell (full translation invariance) can therefore be specified by only a single tensor $A$

\begin{equation}\label{eq:Inf1}
| \psi \rangle = \sum_{s} Tr \left( A^{(s_1)}A^{(s_2)}... \right) | s_1 s_2 ...  \rangle.
\end{equation}

Similarly, a state with two-site translation invariance ($\ell= 2$) specified by two tensors, ${A_1, A_2}$ and has the form

\begin{equation}\label{eq:Inf2} 
| \psi \rangle = \sum_{s} Tr( A_1^{(s_1)}A_2^{(s_2)}A_1^{(s_3)}A_2^{(s_4)}...)| s_1 s_2 s_3 s_4 ... \rangle.
\end{equation}

 Of course, because the sums in Eqs. (\ref{eq:Inf1}) and (\ref{eq:Inf2}) run over an infinite number of sites, in general there is no way to specify or compute the coefficients. Certain expectation values, 
on the other hand, may still be expressed as the limit of an infinite product of transfer matrices. It is quite common, for example, to consider the expectation value of an operator with the same translational invariance as the state in question, by looking at the per-site behavior. For our purposes, we will again be concerned with product operators of the form given in equation \ref{eq:ProdOp}. However, we will now restrict ourselves further by imposing translation invariance on $Q$. For an infinite system with a unit cell of length $\ell$, we shall consider only $Q$ with $Q_j = Q_{j+\ell}$.

In this situation, one can still sensibly define the expectation value as

\begin{equation}\label{eq:InfExpt}
\langle Q \rangle = \frac{1}{\langle \psi | \psi \rangle}Tr \left( \prod_{j = 1}^{\infty}T_\ell \right).
\end{equation}

Here, the transfer matrix $T_\ell$ is now ``enlarged" to represent an entire unit cell of the chain

\begin{equation}\label{eq:TransferProd}
T_\ell \equiv \prod_{j = 1}^{l}T_j.
\end{equation}

In order to approach the infinite case, we shall first examine the case of a finite but very long chain of length $L$, so that the product in Eq. (\ref{eq:InfExpt}) is limited to $L/\ell$ terms, i.e.

\begin{equation}\label{eq:FinInfExpt}
\langle Q \rangle_L = \frac{1}{\langle \psi | \psi \rangle}Tr \left( \prod_{j = 1}^{L/\ell}T_\ell \right).
\end{equation}

For $L$ sufficiently large, the product can then be approximated by considering an eigenvalue decomposition of the transfer matrix $T_\ell = U \Lambda U^{-1}$ and inserting it into Eq. (\ref{eq:FinInfExpt}) (note that, by construction, the transfer matrix as defined by Eqs. (\ref{eq:Transfer}) and (\ref{eq:TransferProd}) is Hermitian and hence diagonalizable). By applying the cyclic property of the trace operation, all $U$ matrices can be made to cancel, leaving us only

\begin{equation}
\langle Q \rangle_L = \frac{1}{\langle \psi | \psi \rangle} Tr \left(\Lambda^{L/\ell} \right),
\end{equation}
or,

\begin{equation}
\langle Q \rangle_L = \frac{1}{\langle \psi | \psi \rangle} \sum_{j = 1}^{\chi^2} \lambda_j^{L/\ell}, 
\end{equation}
where ${\lambda_j}$ are the diagonal elements of $\Lambda$, i.e. the eigenvalues of the matrix . At this point, it can be observed that in the infinite limit, only the largest eigenvalue $\lambda_{max}$ will contribute to the sum. In other words, we have simply

\begin{equation}
\langle Q \rangle_L = \frac{1}{\langle \psi | \psi \rangle} \left( \lambda_{max} \right) ^{L/\ell}. 
\end{equation}

To fix the norm, we consider a particular transfer matrix  $\tilde T_\ell$, defined as usual by Eq. (\ref{eq:Transfer}) for the special case where $Q = \bigotimes_j \mathbb{1}_j$. Then, we calculate $\tilde \lambda_{max}$, the largest eigenvalue of $\tilde T_\ell$, which satisfies

\begin{equation}
\langle \psi | \psi \rangle = \left(\tilde \lambda_{max} \right) ^{L/\ell} .
\end{equation}
Substituting, we have

\begin{equation}\label{eq:Infinite}
\langle Q \rangle_L =  \left( \frac{\lambda_{max}}{\tilde \lambda_{max}} \right)^{L/\ell}. 
\end{equation}
We then gain access to the per-site behavior by means of a logarithm, which gives

\begin{equation}\label{eq:FinPerSite}
 \frac{1}{L} \log \langle Q \rangle_L =  \log \left( \frac {\lambda_{max}}{\tilde \lambda_{max}} ^{1/\ell} \right).
\end{equation}

Eq. (\ref{eq:FinPerSite}), however, does not depend on having a finite $L$. Thus, even for our infinite system, we can consider the limit

\begin{equation}\label{eq:FinPerSite}
\lim_{L \to \infty} \frac{1}{L} \log \langle Q \rangle =  \log \left( \frac {\lambda_{max}}{\tilde \lambda_{max}} ^{1/\ell} \right).
\end{equation}

\begin{figure}[ht]
\includegraphics[width = 90mm]{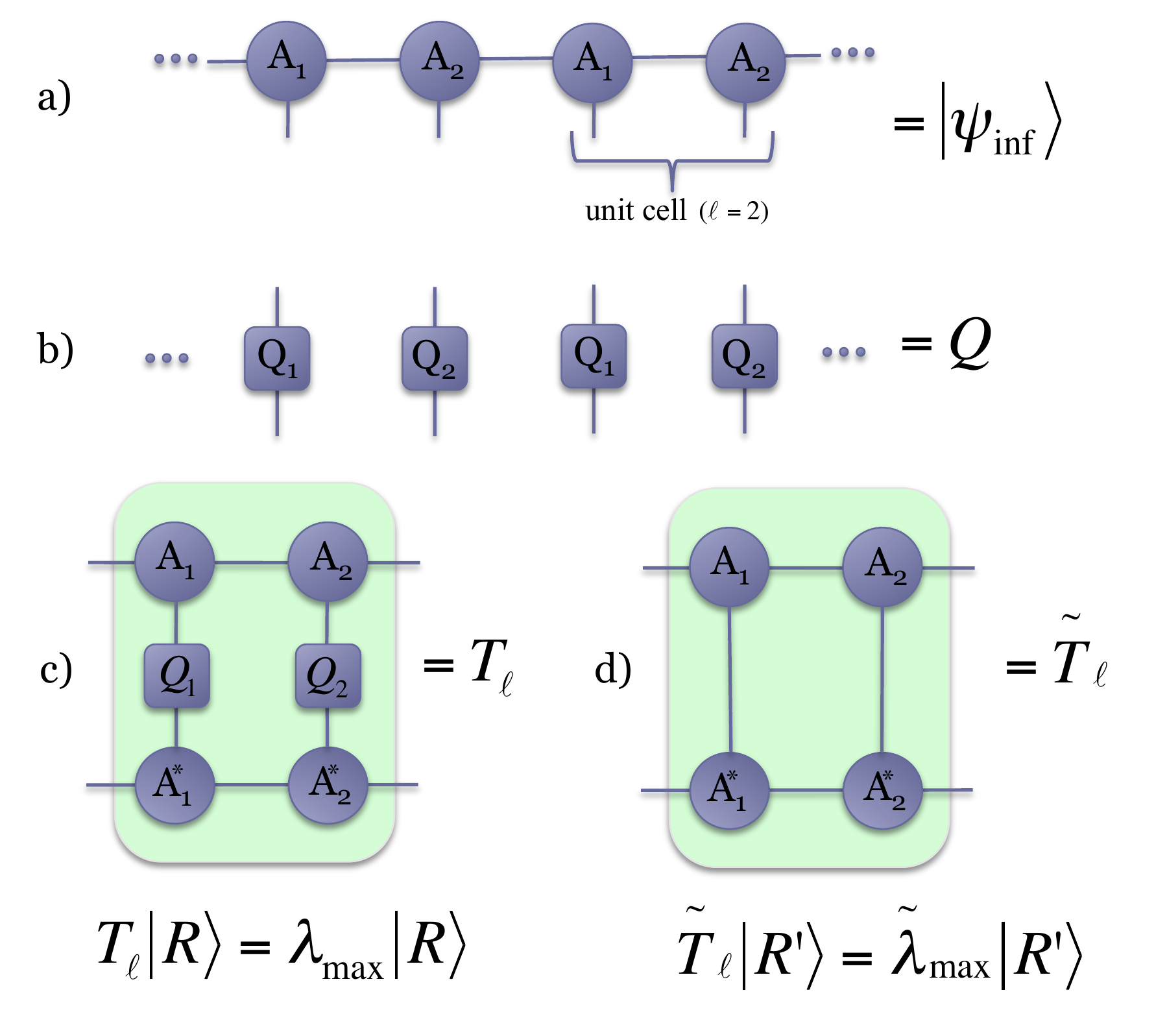}
\caption{\label{fig:MPSInfiniteFig} (a) An infinite spin chain state $| \psi_{inf} \rangle$, possessing translation invariance with respect to a unit cell of length $\ell = 2$, represented as a matrix product state. (b) A product operator $Q = \otimes_j Q_j$ which possesses the same translation symmetry as $| \psi \rangle$; i.e. $Q_j = Q_{j + \ell}$ (c) To compute the quantity of interest, we first construct $T_\ell$, the transfer matrix containing an entire unit cell of $| \psi \rangle$ and $Q$, and extract its dominant eigenvalue $\lambda_{max}$. (d) To normalize the result, we will also need $\tilde T_{\ell}$ (a transfer matrix which contains only the identity operator) and it's dominant eigenvalue $\tilde \lambda_{max}$. The desired quantity $\lim_{L \to \infty} \frac{1}{L} \log \langle Q \rangle$ is given by $ \log ( \lambda_{max}/\tilde\lambda_{max} ) ^{1/\ell}$.}
\end{figure}

Hence, with these procedures (illustrated graphically in Fig.~\ref{fig:MPSInfiniteFig}), we can extract information about product operators in their infinite limit even though their expectation values generally diverge. As we will show below, this information will be sufficient to compute the cumulants of operators with translation invariance even in the infinite case.

\section{Evaluating Higher-Order Moments and Cumulants}
\subsection{Moment and Cumulant-Generating Functions}

For a given operator $M$ and a state $| \psi \rangle$, there is an associated function $F$ which contains all of the non-local information about the higher moments $\langle M^n \rangle$, and yet, as we shall subsequently demonstrate, can still be efficiently evaluated within the framework of a tensor product state. In particular, this function is given by $F(a) \equiv \langle e^{a M} \rangle$. 

In probability theory, $F(a)$ is termed the ``moment generating function" of the probability distribution. It is so named because the information about every moment of the distribution is not only contained, but readily accessible from this single function. This can be made explicit by considering a Taylor-expansion of $e^{a M}$ about $a = 0$ and then computing the expectation value in $F(a)$ term-by-term

\begin{equation}
F(a) = 1 + a \langle M \rangle + \frac{a^2}{2} \langle M^2 \rangle + ...
\end{equation}

From the result, it is clear that every (non-vanishing) moment $ \langle M^n \rangle $ will appear in the expansion. Furthermore, these moments can be directly accessed by computing

\begin{equation}\label{eq:MasterRelation}
F^{(n)}(a) = \mu_n + \mathcal{O}(a),
\end{equation}
where $F^{(n)}(a)$ is as usual the $n^{th}$ derivative of $F$.

The moment-generating function $F$ is closely related to the so-called ``characteristic function" of the distribution, $G(a) \equiv \langle e^{i a M} \rangle$ \cite{CharacteristicFunctions}. For typical states with well-behaved wave functions, these functions will be essentially interchangeable (up to a factor of $i$). Hence in this work, both will be used, sometimes in combination, depending on the particular moment or operator being computed. It should be noted, however, that for some ``pathological" wave functions, such as those specifying a Lorentzian probability distribution, the function $F(a)$ may fail to exist. The characteristic function $G(a)$, however, being the expectation value of a bounded operator, does not suffer from this complication in any situation \cite{Rudin}.

While Eq. (\ref{eq:MasterRelation}) gives a result for the desired moment which is only accurate up to first order in the parameter $a$, the precision can be improved by instead computing appropriate combinations of the functions $F(a)$, $F(-a)$, $G(a)$, and $G(-a)$. For example, when seeking to compute an even-ordered moment; i.e. a moment of the form $\langle M^{2n} \rangle$, we construct

\begin{equation}\label{eq:PlusMinus}
\tilde{F}(a) = F(a)+ F(-a) = 2 + a^2 \langle M^2 \rangle + \frac{a^4}{12} \langle M^4 \rangle + ...
\end{equation}

And hence obtain the desired moments from the relation 

\begin{equation}\label{eq:PlusMinus2}
\tilde{F}^{(n)}(a) \propto \langle M^{2n} \rangle + \mathcal{O}(a^2).
\end{equation}

Odd-ordered moments can of course be found to higher precision from $F(a) - F(-a)$. Even greater precision can also be obtained by including the characteristic functions. The combination $F(a)+ F(-a) -G(a) - G(-a)$, for instance, determines $\langle M^2 \rangle$ up to $\mathcal{O}(a^4)$. 

We employ a similar technique to extract the cumulants of the distribution. This is done by means of the ``cumulant generating function," defined as

\begin{equation}
l_F(a) \equiv \log F(a).
\end{equation}

To see the utility of this function, observe that

\begin{equation}
l_F(a)  \approx \log(1 + a \langle M \rangle + \frac{a^2}{2} \langle M^2 \rangle + ...).
\end{equation}

For small enough values of $a$, one can see from the expansion $\log(1+x) \approx x - \frac{1}{2}{x^2} + ... $ that

\begin{equation}
l_F(a) = a\langle M \rangle + \frac{a^2}{2}\langle M \rangle ^2  + \frac{a^2}{2} \langle M^2 \rangle + \mathcal{O}(a^3).
\end{equation}

Grouping these terms by the powers of $a$, we find that in fact

\begin{equation}
l_F(a) = a \kappa_1 + \frac{a^2}{2}\kappa_2  + ...
\end{equation}

In other words, the derivatives of the function $l_F(a)$ give us direct access to the cumulants in the same manner as the moments in Eq. (\ref{eq:MasterRelation})

\begin{equation}
l_F^{(n)}(a) = \kappa_n+ \mathcal{O}(a).
\end{equation}

Of course, as with the moments, appropriate combinations of $l_F(a)$, $l_F(-a)$ and the associated complex functions can be used to suppress the higher order terms and improve the accuracy. 

\subsection{Evaluating generating functions on a finite matrix product state}

Evaluating all of these moments and cumulants thus boils down to evaluating the expectation values of operators like $e^{aM}$. It remains to be shown that these operators, which we term ``moment generating operators," can be applied in an efficient manner. Fortunately, the exponential structure of the operator guarantees that this is indeed the case. 

We will consider these operators in two cases, depending on the nature of the operator $M$. The first, special case is the large class of operators where $M$ can be written as $\sum_j O_j$, where $j$ runs over all the sites in the system and for some arbitrary set of on-site operators $\{O_j\}$. Most usefully, this set of operators contains the traditional magnetization operators such as $M_x = \sum_j \sigma^x_j$, as well as staggered magnetizations, crystal field magnetizations, etc. Subsequently, we will examine the more general case where the terms within $M$ act on more than one site at a time. 

\subsubsection{Sums of single-body operators}

In this case, since the operators $O_j$ all act at separate sites, the combined operator $M$ is in fact simply a Kronecker sum, $M = \bigoplus_ j O_j$. From this it follows that we can write \cite{Neudecker} 

\begin{equation}\label{eq:Operator}
e^{a M} = \bigotimes_j e^{a O_j}.
\end{equation}

In other words, the moment-generating operator can be decomposed into a set of operators $\{e^{aO_j} \}$, each acting only at a single site. The moment generating function, in turn, is just the expectation value of all these operators applied simultaneously.

Since our moment-generating operator can be written in this tensor product format, we can compute its expectation value directly through Eqs. (\ref{eq:BinderExpt}) and (\ref{eq:Transfer}), with $Q_j = e^{a O_j}$ (see Fig.~\ref{fig:TwoSite}). The calculation of our moment-generating function $F(a)$ has therefore been reduced to the calculation of $L$ transfer matrices and a single trace over their product. In practice, to calculate a higher moment like $\langle M^2 \rangle$, we then need to repeat this procedure and compute $F$ for slightly different values of $a$, so that it is possible to evaluate the necessary derivative numerically. This can be done through any of the wide variety of standard methods; in this work we have used primarily the classic divided difference formulas \cite{Fornberg}. In general, the more values of $a$ at which $F(a)$ is computed, the higher the accuracy of the derivative. However, since the initial $a$ is already chosen to be quite small, in practice it is often the case that only a very small number of points needs to be computed (to check the behavior and accuracy, the procedure can always be repeated with a smaller value of $a$).

Let us examine now the performance of the method for the case of $\langle M^2 \rangle$. A second derivative is necessary, which can be computed to second order in $a$ from three values of $a$, centered at $a = 0$ \cite{Fornberg}. Noting that when $a = 0$ we have trivially $F(a)$ = 1, it follows that we only have to compute two expectation values, each involving the construction of (at most) $L$ transfer matrices, which are then multiplied together and traced over. By comparison, to compute $\langle M^2 \rangle$ directly, as the sum of all correlators $\langle O_j O_k \rangle$, requires the construction of the same number of transfer matrices (to cover the special case of the correlator where $j = k$), but these matrices must be multiplied and traced over up to $L^2$ separate times. Since some of these products of transfer matrices in these calculations will appear more than once, the actual computational cost can be reduced somewhat through use of a suitably ``dynamically programmed" algorithm, where previously calculated products are saved and recycled \cite{MPSReview}. Even in this case, however, far more than two solitary products would be required. Furthermore, as the order of the desired moment $\mu_n$ increases, the advantage of the moment-generating function method becomes increasingly pronounced, as the numerical derivative will require only approximately $n$ expectation values, instead of $L^n$. 

Simply put, the fact that the exponential nature of the moment generating operator turns long Kronecker sums into simple Kronecker products makes it ideally suited for use with a matrix product state. In all cases, only a small number of expectation values must be computed in order to allow the calculation of a numerical derivative, with each expectation value containing the operator $e^{a O_j}$ at every site $j$. Moreover, application of these local operators does not increase the bond dimension of the state. 

Such moments can in principle also be computed by means of an MPO \cite{IanMPO}. As a straightforward demonstration of this, consider for example the MPO given by \cite{Pollmann}

\begin{equation}
\hat{C_j} = 
\left( \begin{array}{cc}
\mathbb{1}_j & 0 \\
O_j & \mathbb{1}_j  \end{array} \right),
\end{equation}
coupled with the boundary conditions 

\begin{equation}
\langle \phi_L |= 
\left( \begin{array}{cc}
0 & 1 \end{array} \right),
\end{equation}
and

\begin{equation}
| \phi_R \rangle= 
\left( \begin{array}{c}
1\\
0 \end{array} \right).
\end{equation}
To evaluate the moments, one defines the total MPO $W$ to be

\begin{align*}
W &\equiv \langle \phi_L | \prod_{j=1}^L \hat{C_j} | \phi_R \rangle \\
&= \sum_{j=1}^L O_j.
\end{align*}
so that

\begin{equation}
\mu_n = \langle \psi | W^n | \psi \rangle.
\end{equation}

In this naive implementation, each application of $W$ to $|\psi \rangle$ increases the bond dimension of the state by a factor of two, and thus, to calculate the $n^{th}$ moment in this way requires a bond dimension exponential in $n$. This increase can be overcome, if necessary, by means of a standard truncation approximation, as done in the TEBD algorithm. Alternatively, a more sophisticated MPO can be constructed \cite{Nebendahl} which represents the operator $M^n$, but with a bond dimension of just $n+1$, resulting in a procedure which still scales linearly.

\subsubsection{Sums of many-body operators}
We now examine the case of calculating the higher-order moments of a more general set of operators $M = \sum_j O_jO_{j+1}...O_{j+k}$. In other words, we consider operators which are a sum of terms acting on at most $k$ sites at a time. So long as $k$ is finite, it remains possible to evaluate the moment generating functions with a single expectation value. This can be done by appealing to the same iTEBD technique  \cite{TEBD, TimeEvolution} widely used to simulate time evolution and imaginary time evolution in tensor network states.

To begin, we partition the terms of the operator into classes of mutually commuting operators. This will require at most $k$ classes. For example, consider as an operator the two-body transverse Ising Hamiltonian

\begin{equation}\label{eq:IsingHamExample}
H = -\sum_j \sigma^x_j \sigma^x_{j+1} + B \sigma^z_j.
\end{equation}

For simplicity, let us write $H$ in a manner which makes it explicitly a sum of two-body operators

\begin{equation}
H = -\sum_j \sigma^x_j \sigma^x_{j+1} + \frac{B}{2} \left( \sigma^z_j + \sigma^z_{j+1} \right).
\end{equation}

Then, the terms can be partitioned between the even and odd pairs of sites, and $H$ can be written as 

\begin{equation}\label{eq:HamPartition}
H = H_{even} + H_{odd},
\end{equation}
with
\begin{equation}
H_{even} = -\sum_{j even} \sigma^x_j \sigma^x_{j+1} + \frac{B}{2} \left( \sigma^z_j + \sigma^z_{j+1} \right)
\end{equation}
and

\begin{equation}
H_{odd} = -\sum_{j odd} \sigma^x_j \sigma^x_{j+1} + \frac{B}{2} \left( \sigma^z_j + \sigma^z_{j+1} \right).
\end{equation}

The moment generating function therefore has the form $F(a) = e^{a H_{even} + a H_{odd}}$, and admits a Suzuki-Trotter approximation \cite{SuzukiOriginal}. To second order in $a$, this has the form

\begin{equation}\label{eq:SuzukiTrotter}
e^{a H_{even} + a H_{odd}} \approx e^{\frac {a}{2} H_{even} } e^{a H_{odd}} e^{\frac{a}{2} H_{even}}.
\end{equation}

Higher order versions of the approximation have also been well-documented and can be easily substituted where greater precision is required \cite{SuzukiTrotter2}. 

\begin{figure}[ht]
\includegraphics[width = 90mm]{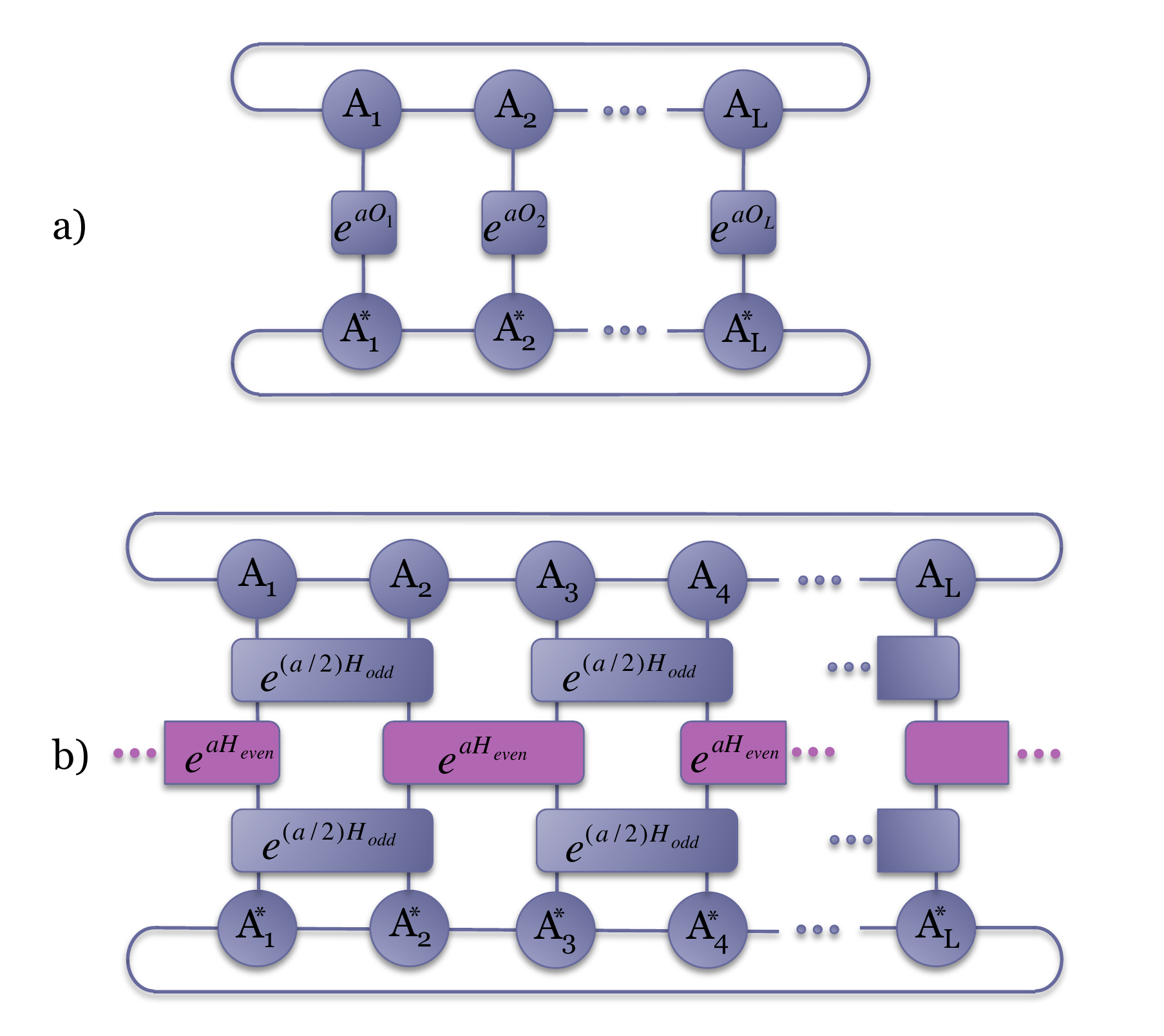}
\caption{\label{fig:TwoSite} (a) Graphical representation of the moment-generating function $F(a) = \langle e^{a O} \rangle$ for an operator $M = \sum_j O_j$. Since each term in $M$ acts at only one site, the moment-generating operator possesses the same structure, even though the moments $M^n$ are fundamentally non-local. (b) The moment-generating function for an operator which is the sum of two-body terms and which possesses the form $H = H_{odd} + H_{even}$, such as the transverse Ising Hamiltonian defined in Eq. (\ref{eq:IsingHamExample}). The operator is approximated by the second-order Suzuki-Trotter formula in Eq. (\ref{eq:SuzukiTrotter}), which produces three ``layers" of operations. Each layer is a sum of two-body terms.
}
\end{figure}

Because $H_{even}$ and $H_{odd}$ were explicitly constructed to be sums of mutually commuting terms, each exponential in the right hand side of Eq. (\ref{eq:SuzukiTrotter}) is now in the same Kronecker sum form as we had in the case of on-site operators, and each can therefore be equivalently expressed as a single tensor product of operations acting on the entire state at once, in the manner of Eq. (\ref{eq:Operator}). This is graphically depicted in Fig.~\ref{fig:TwoSite}. Hence, by applying the three exponentials from Eq. (\ref{eq:SuzukiTrotter}) in sequence, we can easily calculate the expectation value that represents the moment-generating function. 

From this point, evaluating the expectation value is no different than the case of on-site operators. The bulk of the numerical costs are therefore essentially the same for both on-site and many-body operator, with only one difference: applying these layers of exponential operators will increase the bond dimension of the system, and the size of these bonds may need to be ``truncated" by some approximation scheme to keep the system numerically tractable. This, however, is a common practice in the field of MPS algorithms, and is easily done by means of a Schmidt decomposition (see for example \cite{iTEBD}).

 While our example considered an operator with $k = 2$, that is, two-body interactions which could be partitioned into two internally commuting classes, the technique easily generalizes to larger interactions. The Suzuki-Trotter approximations, for example, can be iteratively applied to an operator $e^{A+B+C}$ by first approximating $e^{A+(B+C)}$ in terms of $e^A$ and $e^{B+C}$, and then approximating $e^{B+C}$.

The ability to calculate moments and cumulants for an operator with many-body terms has a particularly useful application in the world of numerical state estimation. A common goal of tensor network algorithms is the calculation of an approximate numerical ground state, from a given Hamiltonian $H$. These algorithms are typically iterative in nature, gradually refining the approximation as the energy $E$ tends towards $E_0$. It is therefore often important to have a means of actively checking this convergence during the course of the algorithm.  The variance (second cumulant) of the Hamiltonian, $\langle \Delta H^2 \rangle = \langle H^2 \rangle - \langle H \rangle ^2$ is well-suited to this task \cite{MPSReview}. For $\epsilon = \sqrt{ \langle \Delta H^2 \rangle}$, there will be an exact eigenvalue $E_{ex}$ within $\epsilon$ of the approximate energy $E$. In other words

\begin{equation}
| E - E_{ex}| \leq \epsilon.
\end{equation}

This quantity can therefore be used as an upper bound on the convergence of the system, and gives a sufficient criterion for halting. Although epsilon is not guaranteed to be small as soon as the energy has converged, if we iterate our algorithms until epsilon $\textit{is}$ very small, we can be assured that the approximate ground state energy is very close to the true value $E_0$.

In \cite{MPSReview}, a dynamically programmed algorithm was given for computing $\langle \Delta H^2 \rangle$ in the context of a finite matrix product state. The method presented here performs at least as efficiently in that case, with the advantage that it can also be applied to infinite (or indeed, higher-dimensional) systems. In \cite{IanMPO}, the same quantity was presented and evaluated by means of an MPO. As discussed above, the MPO technique can in principle be used as an alternative method to compute other moments and cumulants as well, at the cost of allowing the bond dimension to increase.

For the case of the energy cumulant, our method is also particularly well-suited for use with the TEBD/iTEBD algorithms. We have previously remarked that for a Hamiltonian $H$, the calculation of the associated moment-generating operator $e^{a H}$ is essentially identical to an imaginary time-evolution operator with a time step of $\delta t = a$. Each iteration of this algorithm therefore amounts to calculating $e^{a H} | \psi \rangle$, from which the moment-generating function $F(a) = \langle e^{a H} \rangle$ can easily be computed. If one also computes $F(-a) = \langle e^{-a H} \rangle$, the error bound can therefore be computed very efficiently up to order $a^2$ in accordance with Eqs. (\ref{eq:PlusMinus}) and (\ref{eq:PlusMinus2}). 

Performing this check at regular intervals throughout the evolution offers a halting condition to certify convergence of the energy. In some respects, this convergence criterion is superior to the typical methods, which often signal a halt when $\delta E$, the change in the approximate energy between two successive iteration steps, drops below some minimum value. Such a method can occasionally give a false sense of convergence when the algorithm ``stalls out" and begins evolving only very slowly, despite remaining some distance from the ground state. The variance of the energy provides information not about the convergence of the algorithm but of the energy itself, by identifying when the system is very close to an exact eigenstate. However, in some cases care must be taken that the nearby eigenstate is in fact the ground state, and not some excitation. 

\subsection{Evaluating generating functions on an infinite matrix product state}
 
At first glance, it may seem that moment-generating techniques discussed above cannot be applied to these systems, since the value of a quantity like $M = \sum_j^{\infty} O_j$ is clearly diverging, and only related limits like 

\begin{equation}
\langle M \rangle = \lim_{L \to \infty} \frac{1}{L} \sum_j^\infty \langle O_j \rangle
\end{equation}
are well-defined. In this situation, however, while the moment-generating function $F$ defined above may diverge, one can still define and calculate the related quantity

\begin{equation}
F_{\infty} = \lim_{L \to \infty} \langle e^{a M} \rangle ^{1/L}.
\end{equation}

As discussed in section III, quantities of this form can in fact be computed quite naturally. Using equation \ref{eq:Infinite} (see Fig.~\ref{fig:MPSInfiniteFig}), clearly we have

\begin{equation}
F_{\infty} = \lim_{L \to \infty} \langle e^{a M} \rangle^{1/L} =  \frac{\lambda_{max}}{\tilde \lambda_{max}}.
\end{equation} 

In other words, the desired quantity is simply the largest eigenvalue of the transfer matrix associated with the state and operator in question. Hence by defining

\begin{equation}
 l_{F_{\infty}}(a) = \log  F_{\infty}(a), 
 \end{equation}
we find that we have access to the per-site limits of the cumulants even in the infinite case. In the same manner as with finite systems, they are given by the derivatives of $l_{F_\infty}$ with respect to $a$ 

\begin{equation}
l_{F_{\infty}}^{(n)}(a) = \lim_{L \to \infty} \frac {1}{L} \kappa_n+ \mathcal{O}(a).
\end{equation}

We note briefly some practical considerations that are important when evaluating $l_{F_{\infty}}$ for a real matrix product state. First, algorithms for generating the states which are based on the iTEBD principle are likely to require the use of a two-site unit cell, even if the state is expected to possess only one-site translation invariance, as a result of the two-body nature of most parent Hamiltonians. In this case, of course, a two-site transfer matrix is required ($\ell = 2$), and we must take a square root of its largest eigenvalue in order to recover the correct per-site limit. 

Additionally, we observe that for the second cumulant in particular, it can be particularly desirable to calculate using the characteristic function $G = \langle e^{i a M} \rangle$ instead of $F$. This is because one can then take advantage of the fact that

\begin{align*}
\lim_{L \to \infty} \frac {1}{L} \kappa_2 &= l_{G_{\infty}}(a) + l_{G_{\infty}}(-a) + \mathcal{O}(a^2)\\
&= \log G_{\infty}(a)+  \log G_{\infty}(-a) 
\end{align*}

Then, combining the two log terms and observing that $G_{\infty}(-a) = G_{\infty}(a)^*$, we have   

\begin{equation}
\lim_{L \to \infty} \frac {1}{L} \kappa_2 = \log \left( |G_{\infty}(a)|^2 \right) + \mathcal{O}(a^2).
\end{equation}

In other words, we can calculate the per-site second cumulant up to second order in $a$ by evaluating $G_{\infty}(a)$ only once, and without directly computing any numerical derivative. 

\subsection{Higher-Dimensional States}
\begin{figure}[ht]
\includegraphics[width = 90mm]{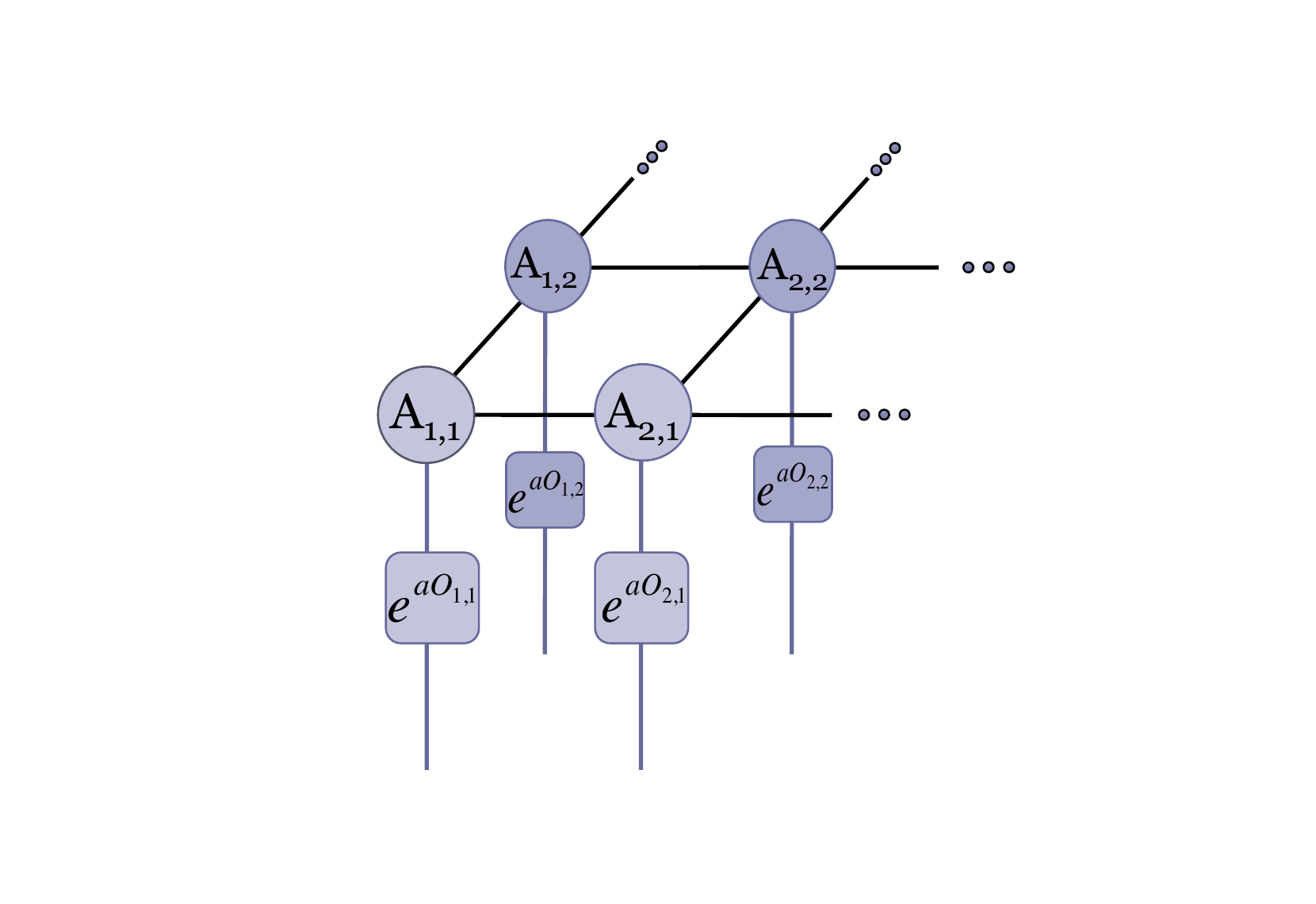}
\caption{\label{fig:2DBinder} The moment-generating operator $e^{aM}$ for an operator of the form $M = \sum_j \sum_k O_{jk}$, applied to a two-dimensional state on a square lattice. As in the one-dimensional case, the locality of each term in $M$ ensures the locality of the terms in $e^{aM}$, and hence, the moment-generating operator can still be evaluated all at once, by applying the appropriate onsite operator at each lattice site.
}
\end{figure}

Although in this work we shall be applying these techniques to one-dimensional systems, there is nothing about the procedures above that cannot be immediately generalized to finite-sized higher dimensional states. Consider for example a total magnetization-type operator on a 2 dimensional, $L \times L$ square lattice, given by

\begin{equation}
M = \sum_j \sum_k O_{jk},
\end{equation}
where$O_{jk}$ represents a specific operator acting locally on site $(j, k)$ of the lattice. Of course by representing both $j$ and $k$ by some composite index $J$ (now running from 1 to $L^2$) we can immediately see that $M$ is no different than the magnetization-like operators we considered in the one-dimensional case

\begin{equation}
M = \sum_J O_J,
\end{equation}
and hence that our earlier analysis goes through: the moment generating operator $e^{a M}$ can still be written as

\begin{equation}
e^{a M} = \bigotimes_J e^{aO_J}.
\end{equation}
This object is still an operator that acts only locally and whose expectation value can be evaluated all at once (graphically depicted in Fig.~\ref{fig:2DBinder}).

If we consider instead an operator $H$ which contains many-body terms (but for whom each term acts nontrivially only on a finite number of sites), one can play the same tricks as in one dimension, first partitioning the terms into mutually commuting sets of terms $H_1, H_2 + H_3...$, then expressing the moment-generating operator as

\begin{equation}
 e^{aH} = e^{a(H_1+H_2 + H_3...)},
 \end{equation}
and finally applying some form of Suzuki-Trotter approximation as described above to express the operator as a product of exponentials, each of which can be applied to the state all at once. For nearest-neighbor interaction terms on a square lattice, the procedure is essentially identical to the one dimensional case, except that one must use four classes instead of two: two for interactions in the horizontal direction, and two for the vertical. For operators with more complicated terms (such as a sum of ``plaquettes,") the number of partitions may be larger, but in general the costs do not grow rapidly despite the increase in system dimension. 

Hence, in either case, the associated moment-generating operators can be disentangled into tensor products or sequences of tensor products, even in higher dimensions. Simply put, the essential ``power" of the moment-generating function method is the fact that the moment-generating operator $e^{aM}$ of a local operator $M$ is itself a local operator, and this fact is unchanged regardless of the dimensionality of the system.

Once the moment-generating operator has been expressed as a tensor product and applied to the state, it still remains to numerically contract the tensor network. It is at this stage where things become more difficult than in the one-dimensional case, since computing the expectation value of $\textit{any}$ operator on a higher-dimensional tensor network state can be quite hard. Exact calculation has been shown to be exponentially costly in $L$ (in particular, it is a $\#$P-hard problem \cite{PepsComplexity}). Nevertheless, a wide variety of numerical techniques  have been developed to approximate these contractions efficiently with minimal errors, the details of which are outside the scope of this paper (See for example refs. \cite{PEPSContract1, PEPSContract2, PEPSContract3, PEPSContract4, PEPSContract5, PEPSContract6}). We do caution, however, that in our experience, the higher the order of the moment, the higher the sensitivity of the result to the errors introduced by approximate contraction.

\section{Examples}
\subsection{Spin-1/2 Transverse Ising Model}
As has become almost customary, we begin by demonstrating our technique in the context of the widely-studied transverse Ising model; a chain of spin-1/2 particles governed by the Hamiltonian

\begin{equation}\label{eq:IsingHam}
H = -\sum_j^L \sigma_j^x \sigma_{j+1}^x + B \sigma_j^z.
\end{equation}

This model is a useful proving ground as it has been extensively studied and admits a well-known analytical solution \cite{IsingOriginal, Ising}, as well as possessing a straightforward order parameter of $M_x = \sum_j \sigma_j^x$, the total magnetization in the $x$-direction. We can therefore test our techniques by using them to study the phase transition known to occur exactly at $B_c = 1$. To apply the Binder cumulant technique, we first use a numerical method to find the ground state by solving the generalized eigenvalue problem \cite{MPSReview}. We consider system lengths between $L=10$ and $L=45$ in steps of five, using a bond dimension of $\chi = 10$ (we verify that increasing the bond dimension does not change the results of the methods up to our working precision).  For each system length, ground states of the Hamiltonian in Eq. (\ref{eq:IsingHam}) are computed as we sweep over a range of values for the field coefficient $B$. Then for each value of $L$ and $B$, we compute the Binder cumulant using the methods described above, by first computing $\mu_2$ and $\mu_4$ by means of the moment-generating function. 

\begin{figure}[b]
\includegraphics[width = 90mm]{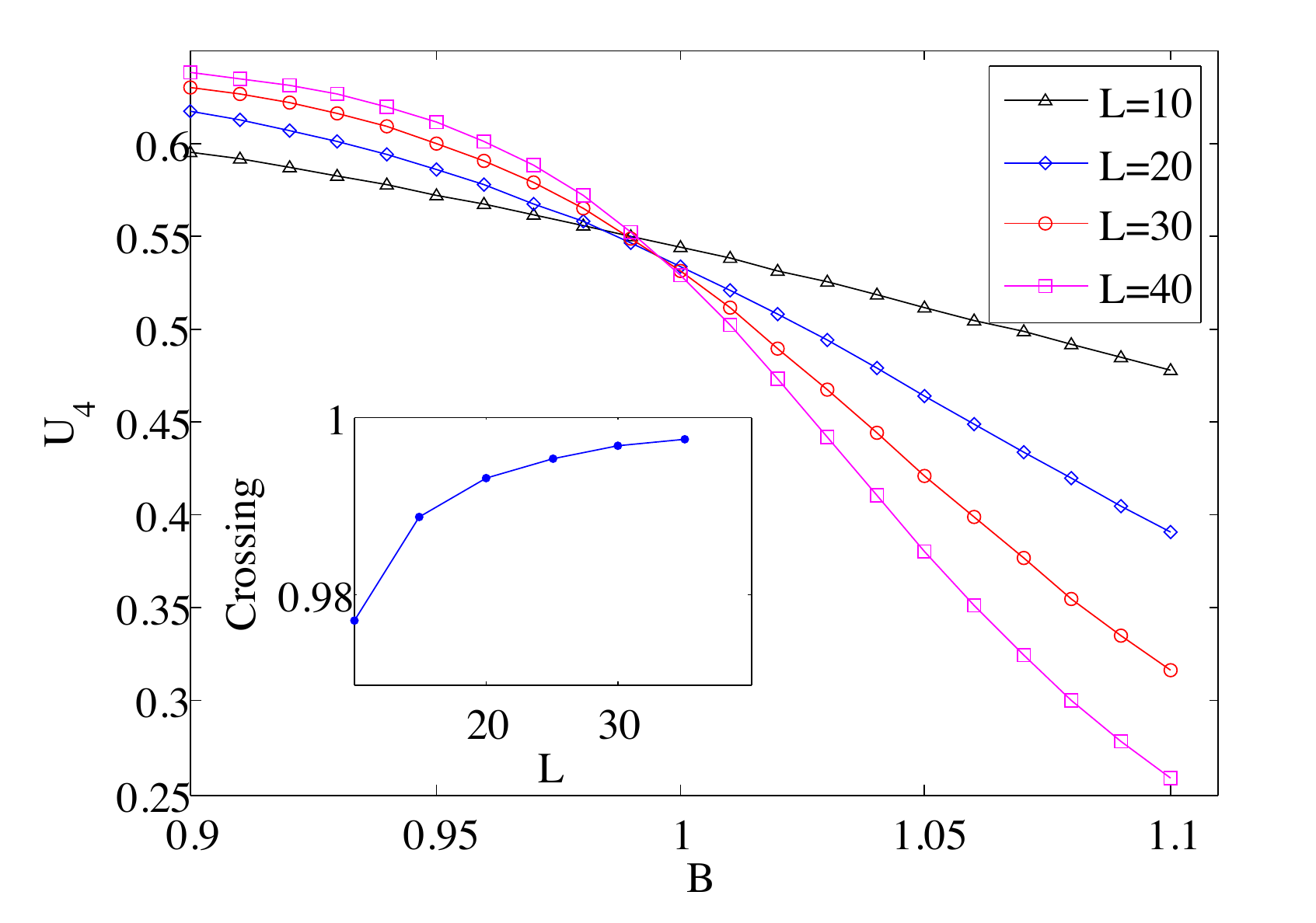}
\caption{\label{fig:IsingCrossings} (color online) A Binder cumulant study of the transverse Ising model. The cumulants are computed for different system sizes across a range of values for the transverse field $B$ (some intermediate system sizes have been suppressed for clarity of the figure). Crossing points are interpolated for successive pairs of curves, i.e. L = 10 and L = 15. These crossing values can then be seen to approach the known value of the critical field, $B_c = 1$ (inset). The BST algorithm is used to extrapolate these values to the infinite limit, which gives $B_c = 1.001(1)$. }
\end{figure}

As shown in Fig.~\ref{fig:IsingCrossings}, the crossings of the Binder cumulants at various lengths are already clustered very close to the transition point, even though the lengths of the states are relatively short compared to the thermodynamic limit. But the location of the critical point can be computed to even greater accuracy by considering the pattern of successive crossings. These crossings show a clear trend towards a limiting value as the system sizes increase. This limiting value can be estimated by means of the BST Algorithm \cite{BSTOriginal}, which has been found to be a very powerful tool for estimating the infinite limit of a series of data based on finite size corrections which obey a power-law, even from a relatively small number of data points \cite{BSTLattice}. From this extrapolation, we estimate a critical point of $Bc = 1.001(1)$. Here and elsewhere, the reported uncertainty in our extrapolation represent an estimate of the typical effect of uncertainties in the location of the crossing points, propagated through the BST Algorithm. Much more detail about the BST technique can be found in Appendix A.

open 

\begin{figure}[ht]
\includegraphics[width = 90mm]{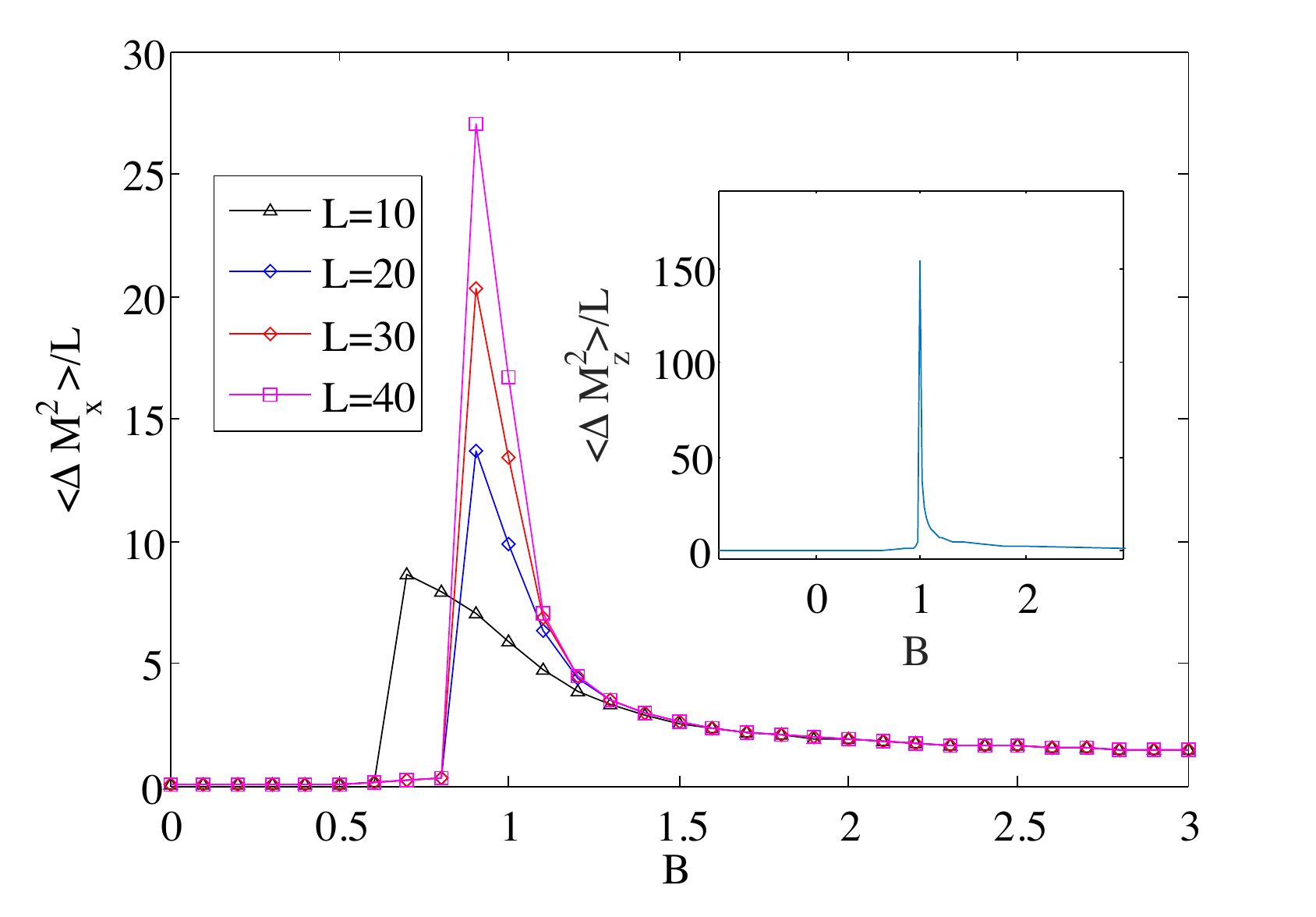}
\caption{\label{fig:IsingM2x} (Color online) Per-site value of the second cumulant of the longitudinal magnetization, $\frac{1}{L} \langle \Delta M_x^2 \rangle = \frac{1}{L} (\langle M_x^2 \rangle - \langle M_x \rangle^2)$, for the transverse Ising model. The cumulant is plotted for various finite system sizes, plotted against a range of applied fields. As the system length increases, the behavior tends towards the infinite limit (inset). In the limit, the cumulant diverges at the critical point. }
\end{figure}

The critical point of the Ising model can be probed directly through the higher-order cumulants of the order parameter, as well. Through the techniques above, these can be calculated from the finite systems at various system sizes. Alternatively, we can calculate a ground state for the infinite system through the iTEBD algorithm (in this case using a bond dimension of $\chi = 20$) and then calculating the second cumulant directly. Both procedures are showcased in Fig.~\ref{fig:IsingM2x}. The behavior in the infinite case can be seen to agree with the limiting trend of the finite systems as the length is increased. The cumulant can be seen to become singular near the critical point, and can also be used to detect the transition. Using this method, we estimate $B_c = 1.00(1)$

\begin{figure}[ht]
\includegraphics[width = 90mm]{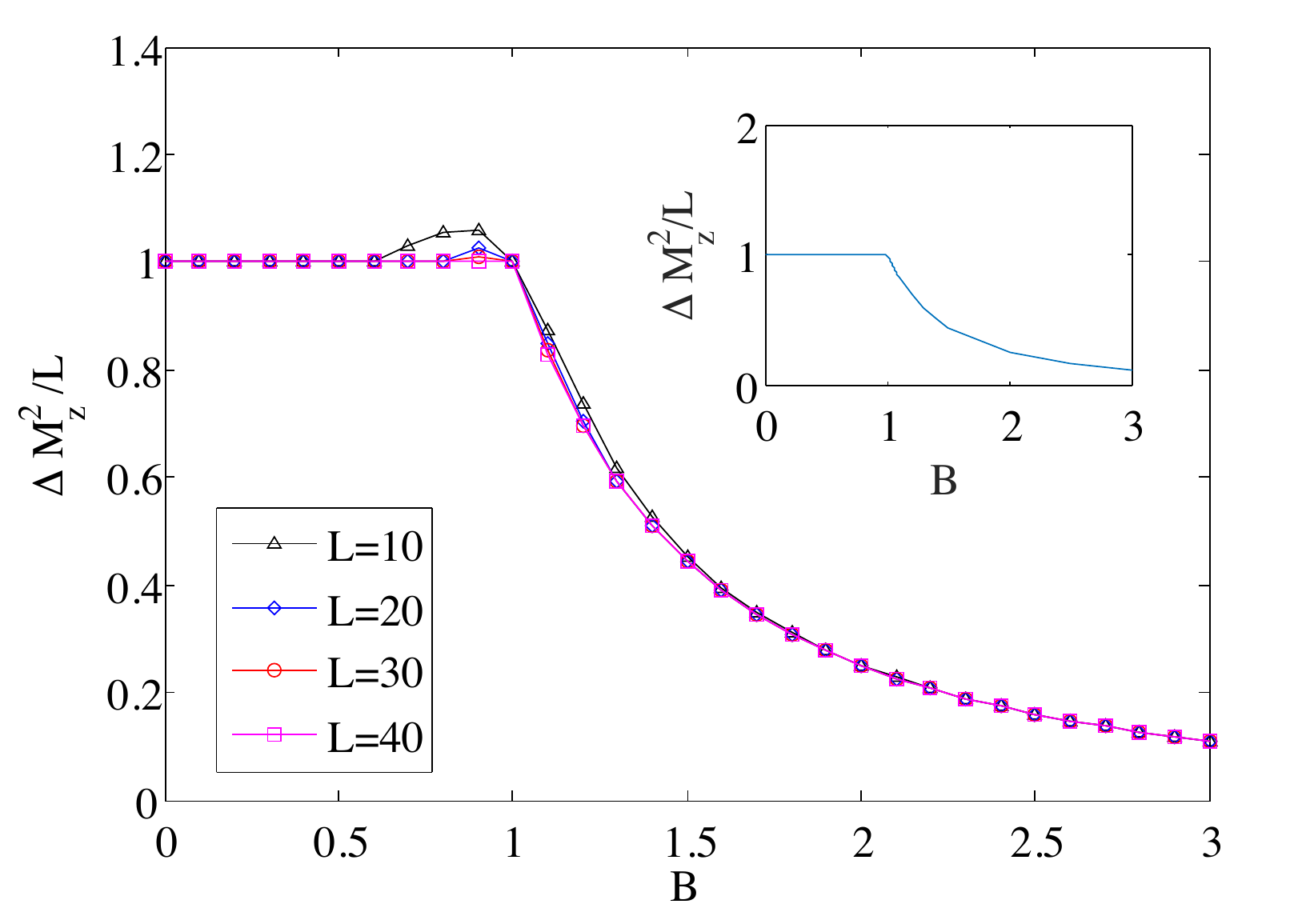}
\caption{\label{fig:IsingM2z} (Color online) Second cumulant of the transverse magnetization, $\langle \Delta M_z^2 \rangle= \langle M_z^2 \rangle - \langle M_z \rangle^2$, for the transverse Ising model (computed per site). The cumulant is plotted for various finite system sizes, plotted against a range of applied fields. As the system length increases, the behavior tends towards the infinite limit (inset). In the limit, the derivative of the cumulant shows a discontinuity at the critical point.}
\end{figure}

This technique can also be applied to the magnetization in the transverse direction, $M_z = \sum_j \sigma_j^z$. In this case, it is the derivative of the cumulant which becomes singular in the thermodynamic limit to signify the critical point. Again, the results from the finite chains can be seen trending towards the infinite limit (see Fig.~\ref{fig:IsingM2z}). 

\begin{figure}[ht]
\includegraphics[width = 90mm]{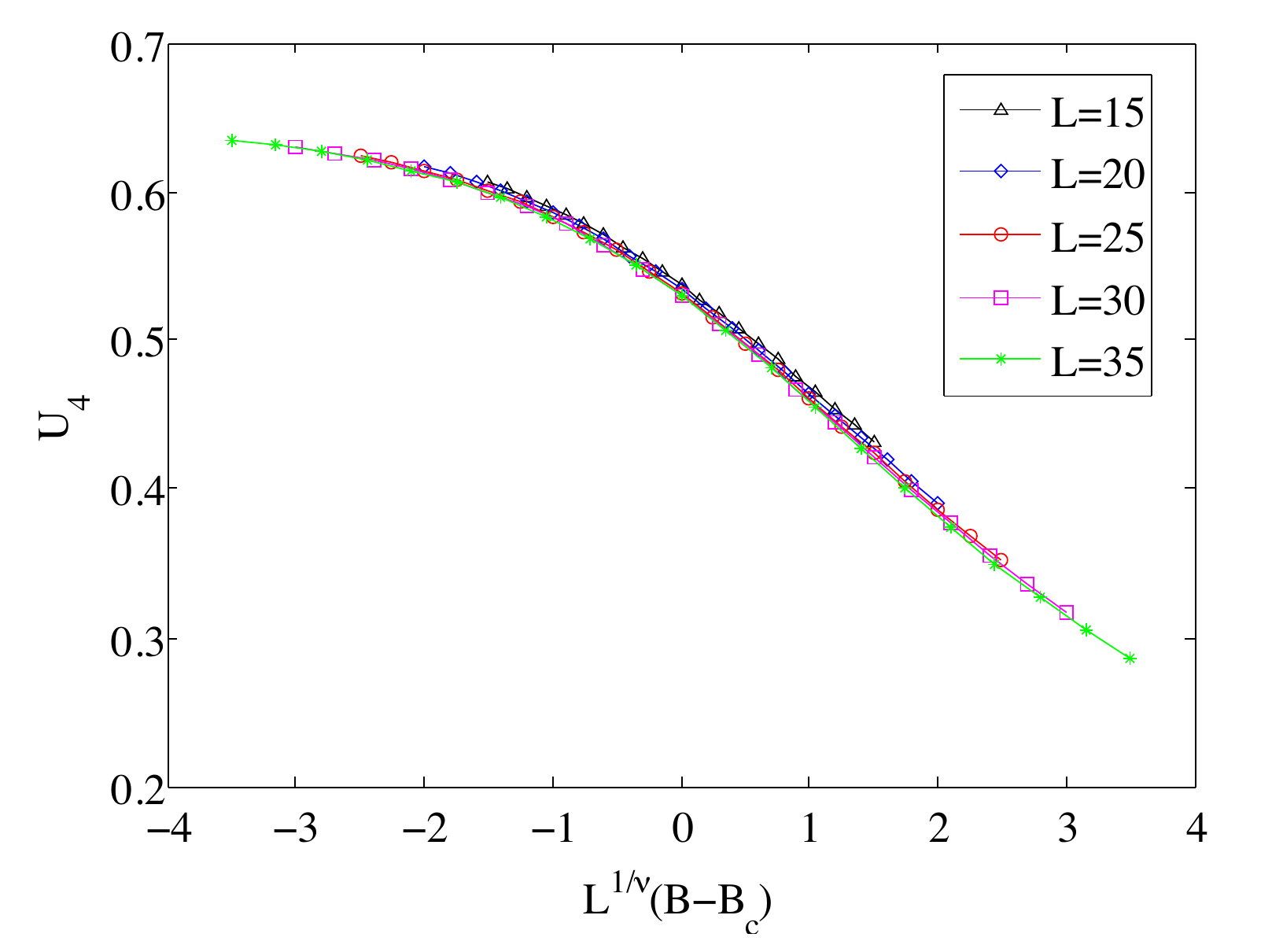}
\caption{\label{fig:IsingDataCollapse} (Color online) The Binder cumulants for the transverse Ising model, plotted for a variety of system sizes as a function of $L^{1/\nu}(B-B_c )$ for the known values $\nu = 1$ and $B_c = 1$. As expected, for these values the curves are seen to collapse to a functional form essentially independent of the length scale. This property can be used to estimate the values of the critical point and the critical exponent by treating them as fit parameters and optimizing the collapse.} 
\end{figure}

The critical exponent of the correlation length of the model can also be studied by means of the Binder cumulant. Once the critical point has been estimated, the curves of $U_4$ can then be plotted against $L^{1/\nu}(B-B_c)$ for various values of $\nu$. At the true critical exponent, the data should ``collapse" to a single functional form independent of $L$, as seen in Fig.~\ref{fig:IsingDataCollapse}.

\begin{figure*}[ht]
\includegraphics[width = 17cm]{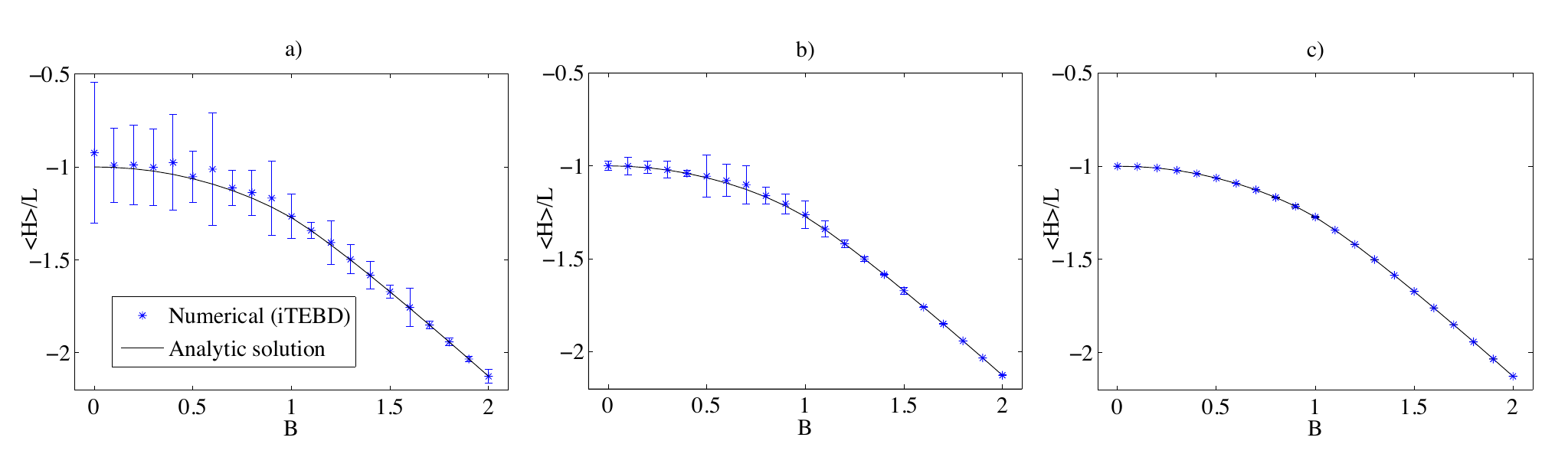}
\caption{\label{fig:H2Combined} (Color online) The energy of the spin-1/2 transverse Ising model is calculated using approximate ground states generated by the iTEBD algorithm ($\chi = 20$). The numerical data (points) are plotted alongside the exact solution (line). Error bars are calculated from $\epsilon = \sqrt{\langle \Delta H^2 \rangle}$, with $\langle \Delta H^2 \rangle$ the second cumulant of the energies. (a) After 10 steps, the energies are still noisy and the error bars are quite large. (b) After 20 steps, the error bars have clearly decreased, and are largest for the points with the largest discrepancies from the exact solution. (c) By 100 steps, the error bars are within the size of the data points, and the approximate energies are very close to the known analytical result.}
\end{figure*}

Finally, we can use this model to demonstrate the utility of the energy variance $\langle \Delta H^2 \rangle$ in assessing numerical convergence, as described above. Starting from a random state, we apply the iTEBD algorithm with the transverse Ising Hamiltonian and evolve towards the ground state, over a range of field strengths $B$. As shown in Fig.~\ref{fig:H2Combined}, the results are initially somewhat noisy when compared to the analytically known $E$ vs. $B$ curve, which is reflected by the large error bars computed from $\langle \Delta H^2 \rangle$. However, as the algorithm continues, these error bars shrink and eventually become essentially zero, signaling the complete convergence of the energies. 

\subsection{Spin-1 Transverse Ising Model}
We consider next the spin-1 generalization of the Ising model, with Hamiltonian

 \begin{equation}\label{eq:Spin1Ham}
H = -\sum_j^L S_j^x S_{j+1}^x + B S_j^z.
\end{equation}

Here, we have simply replaced the spin-1/2 Pauli matrices from Eq. (\ref{eq:IsingHam}) with their spin-1 counterparts. This model is of interest because unlike the spin-1/2 case, it has no exact analytic solution. Nevertheless, in the thermodynamic limit the magnetization is qualitatively similar to the spin-1/2 case. Notably, it still displays a quantum phase transition at a critical value of the transverse field, which has been studied by various numerical \cite{Spin11, Spin12}. The accepted value for this critical field is $B_c = 1.326$ \cite{Spin11}.

\begin{figure}[ht]
\includegraphics[width = 90mm]{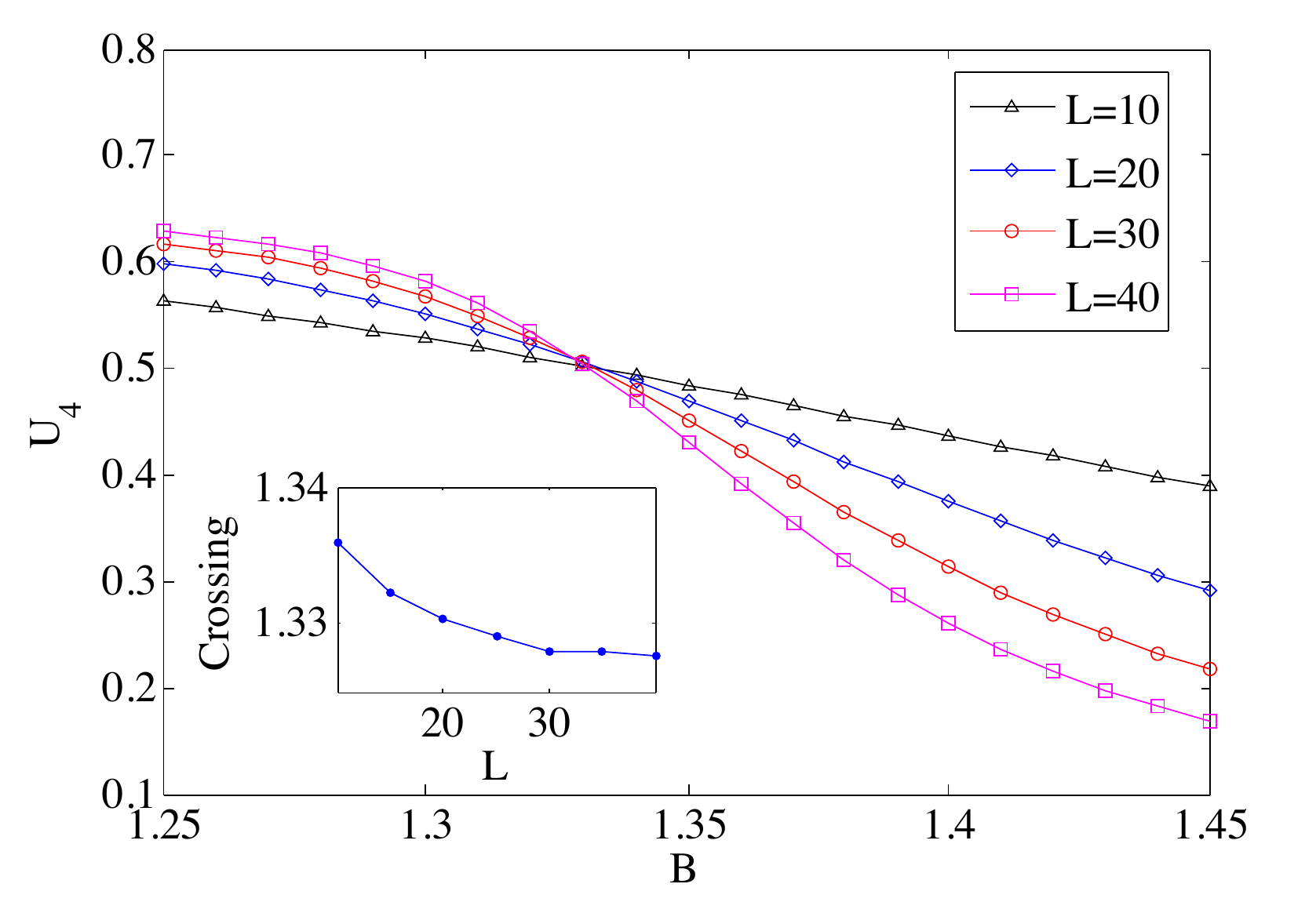}
\caption{\label{fig:BinderSpin1Fig} (Color online) A Binder cumulant study of the spin-1 transverse Ising model. As above, the cumulants are computed for different system sizes across a range of values for the transverse field $B$ (some intermediate system sizes have been suppressed for clarity of the figure). Crossing points are interpolated for successive pairs of curves, i.e. L = 10 and L = 15, and the BST algorithm is used to extrapolate these values to the infinite limit, which gives $B_c = 1.327(1)$. }
\end{figure}

We study this transition point with the same techniques as before, calculating the Binder cumulants from the second and fourth moments of the $x$-magnetization for various system sizes (Fig.~\ref{fig:BinderSpin1Fig}). The ground states are calculated using the same finite MPS technique and with a bond dimension of $\chi = 20$. The successive crossings are then compared and a limiting value extrapolated using BTS. In this case, we compute an estimate of the transition at $B_c = 1.327(1)$.

\begin{figure}[ht]
\includegraphics[width = 90mm]{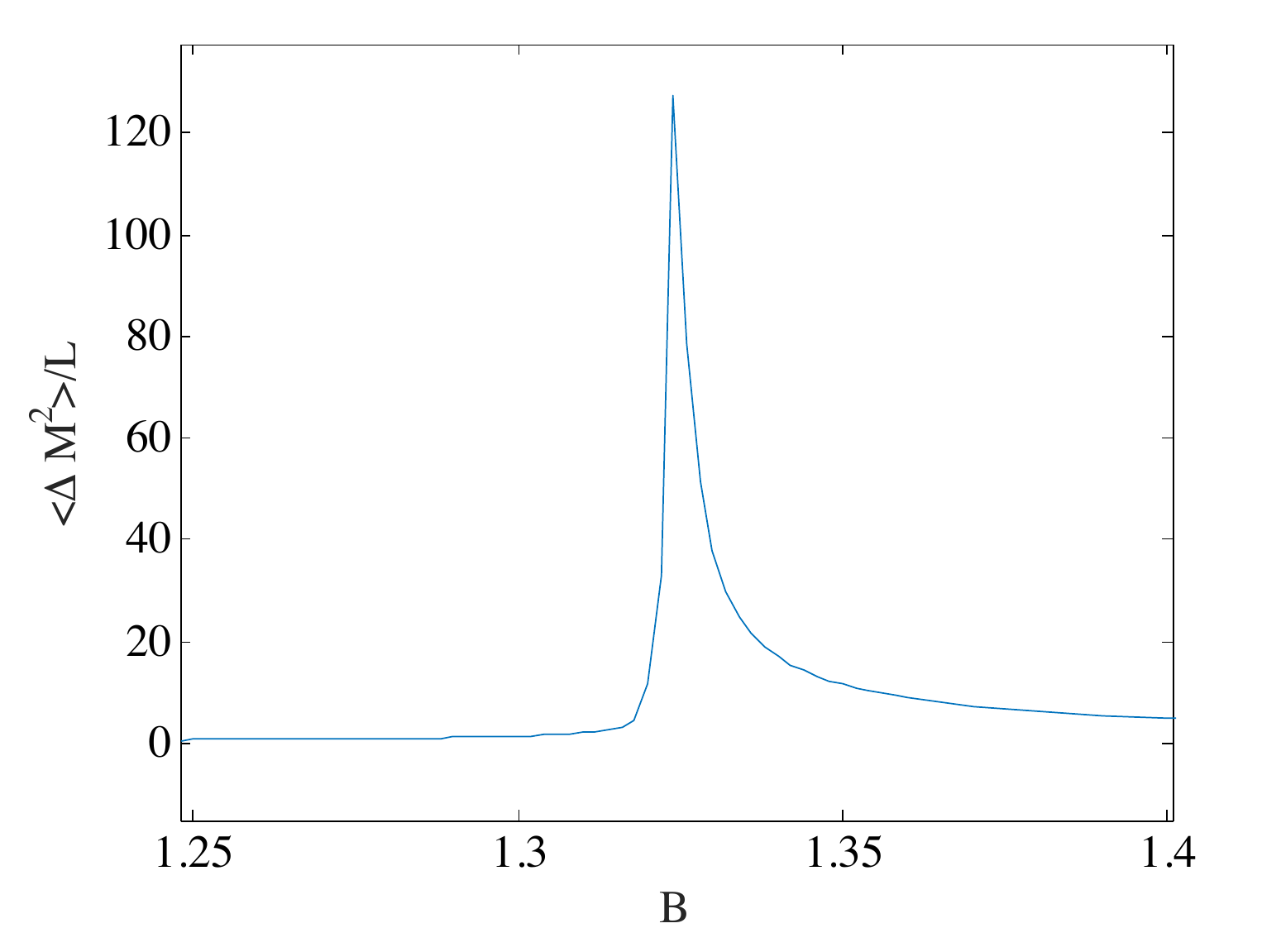}
\caption{\label{fig:Spin1M2x} Per-site value of the second cumulant of the longitudinal magnetization, $\frac{1}{L} \langle \Delta M_x^2 \rangle = \frac{1}{L} (\langle M_x^2 \rangle - \langle M_x \rangle^2)$, computed for the Spin-1 Ising model. The cumulant is calculated for an infinite system directly.}
\end{figure}

Once again, it is also constructive to consider the second cumulant on its own. Numerical calculations of the magnetization for this model invariably show some finite size effects around the transition, producing a finite ``tail" near the transition point, which makes an exact determination difficult using the order parameter alone. But the transition appears much more sharply as a singularity when we consider the second cumulant, as in Fig.~\ref{fig:Spin1M2x} (higher cumulants such as $\kappa_4$ can also be used for this purpose). From this quantity, we obtain $B_c = 1.324(2)$,  an estimation which is to within less than $0.2\%$. 

\begin{figure}[ht]
\includegraphics[width = 90mm]{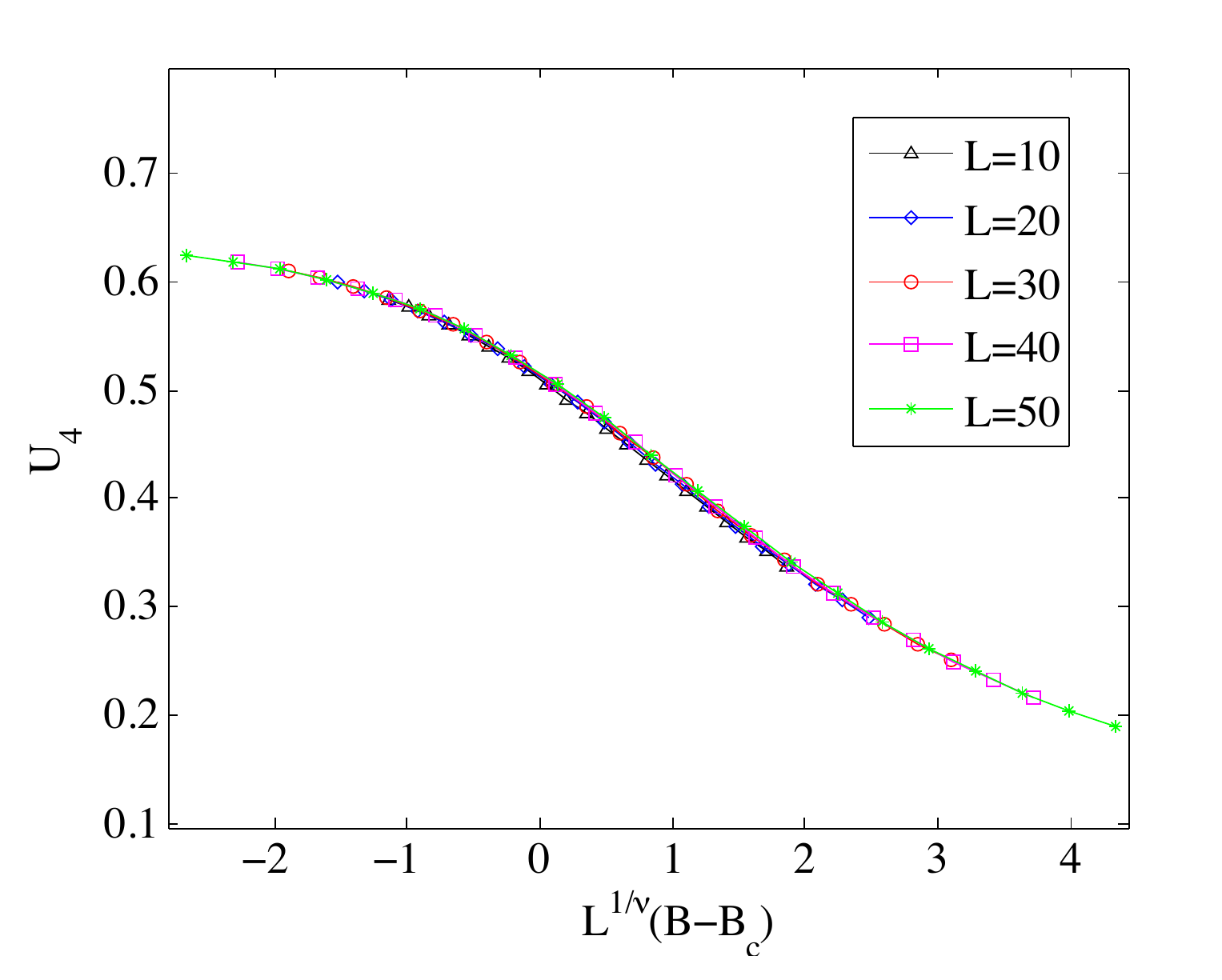}
\caption{\label{fig:Spin1Collapse} (Color online) The Binder cumulants for the spin-1 transverse Ising model, plotted for a variety of system sizes as a function of $L^{1/\nu}(B-B_c )$ for the known values $\nu = 1$ and $B_c = 1.326$. The length-independence of the curves allows this technique to be used as a means to estimate both $\nu$ and $B_c$} 
\end{figure}

As before, can also study the critical exponent $\nu$, known for this model to be the same as the spin-1/2 case, $\nu = 1$. As a proof of principle, the ``data collapse" for the known values of $\nu$ and $B_c$ are shown in Fig. \ref{fig:Spin1Collapse}. As expected, for these values the curves are seen to collapse to a functional form essentially independent of the length scale. This property can be used to estimate the values of the critical point and the critical exponent by treating them as fit parameters and optimizing the collapse.

\subsection{Spin-1 Ising Model in Crystal Field}
For another application, we consider also a variation on the spin-1 Ising model, where the usual transverse field has been replaced by a quadratic, crystal field term, to give the following Hamiltonian

\begin{equation}
H = -\sum_j^L S_j^x S_{j+1}^x + B (S_j^z)^2.
\end{equation}

\begin{figure}[ht]
\includegraphics[width = 90mm]{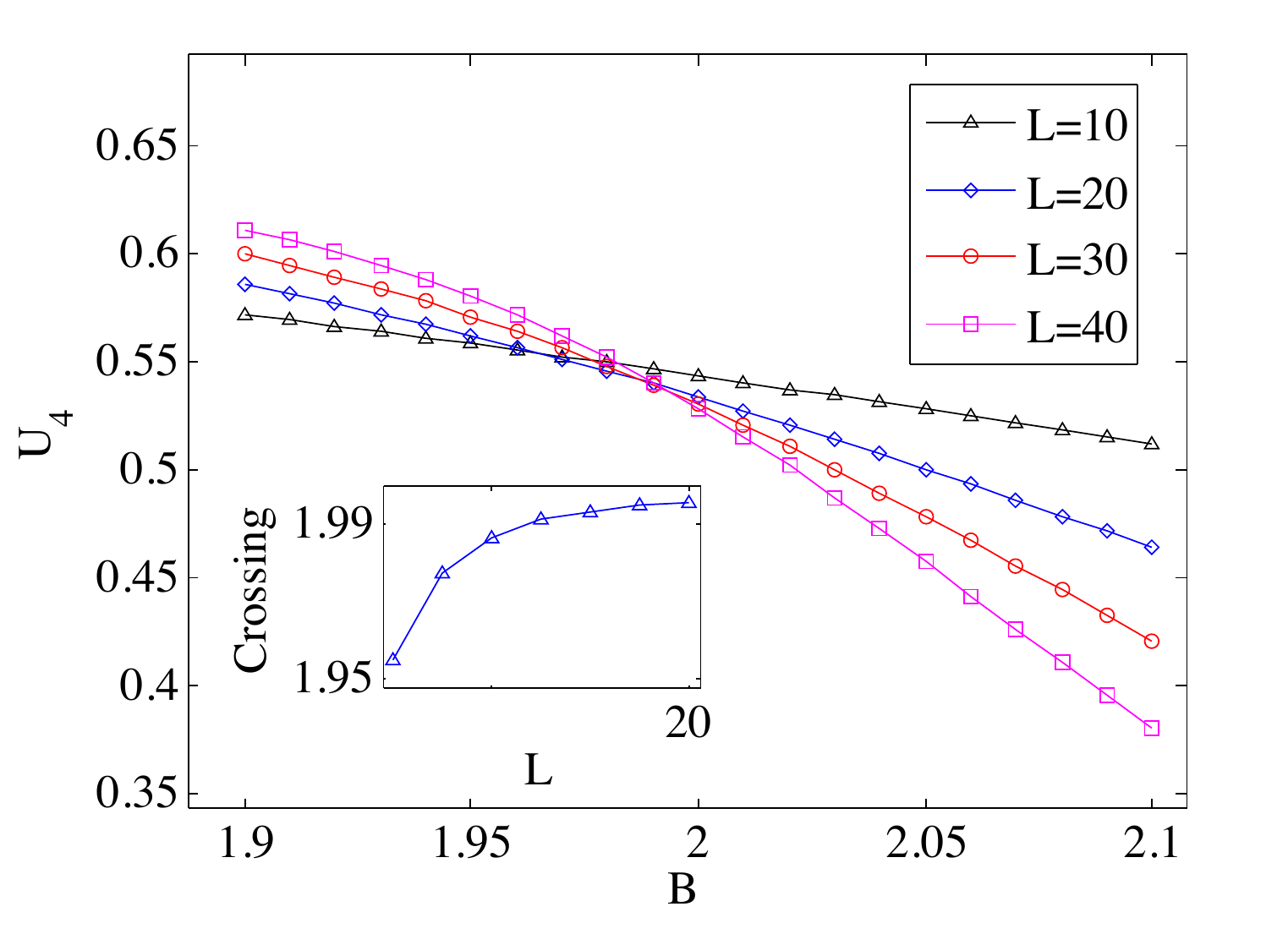}
\caption{\label{fig:BinderCrystalFig} (Color online) A Binder cumulant study of the spin-1 transverse Ising model with crystal field. As above, the cumulants are computed for different system sizes across a range of values for the transverse field $B$ (some intermediate system sizes have been suppressed for clarity of the figure). Crossing points are interpolated for successive pairs of curves, i.e. L = 10 and L = 15, and the BST algorithm is used to extrapolate these values to the infinite limit, which gives $B_c = 1.999(1)$. }
\end{figure}

This variation of the spin-1 model admits a mapping to the spin-1/2 case (c.f. \cite{Crystal}), from which the critical point of $B_c = 2$ can be analytically obtained. To compare our method, we perform the same numerical calculations as above: first generating ground states at various finite lengths using MPS methods with a bond dimension of $\chi = 10$, and then computing the Binder cumulants. From the Binder curves (see Fig~\ref{fig:BinderCrystalFig}) we once again perform the BST extrapolation of the successive crossings to arrive at an estimate of $B_c = 1.999(1)$. 

\begin{figure}[ht]
\includegraphics[width = 90mm]{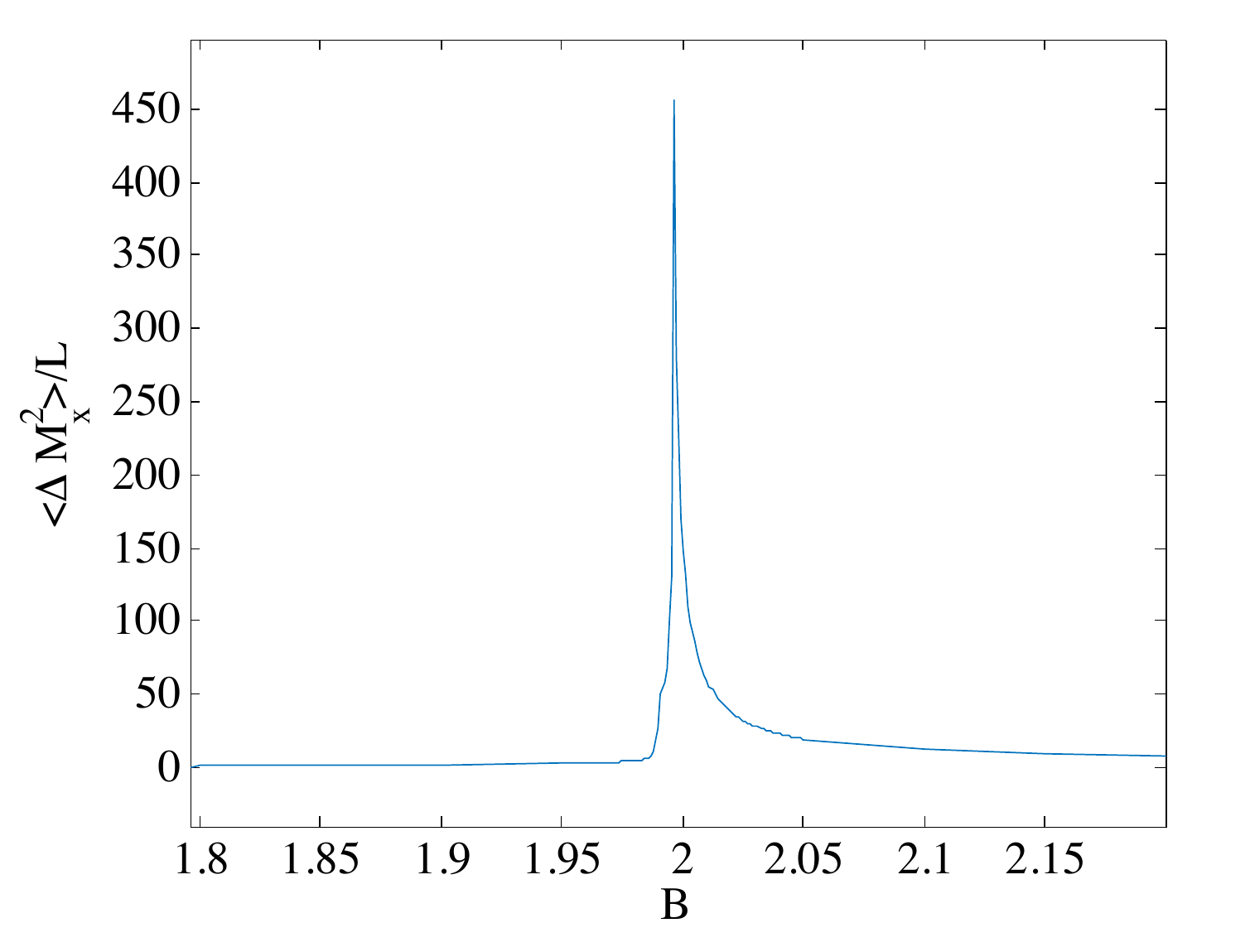}
\caption{\label{fig:CrystalM2x}  Per-site value of the second cumulant of the longitudinal magnetization, $\frac{1}{L} \langle \Delta M_x^2 \rangle = \frac{1}{L} (\langle M_x^2 \rangle - \langle M_x \rangle^2)$, computed for the spin-1 Ising model with a transverse crystal field. The cumulant is calculated for an infinite system.}
\end{figure}

\begin{figure}[ht]
\includegraphics[width = 90mm]{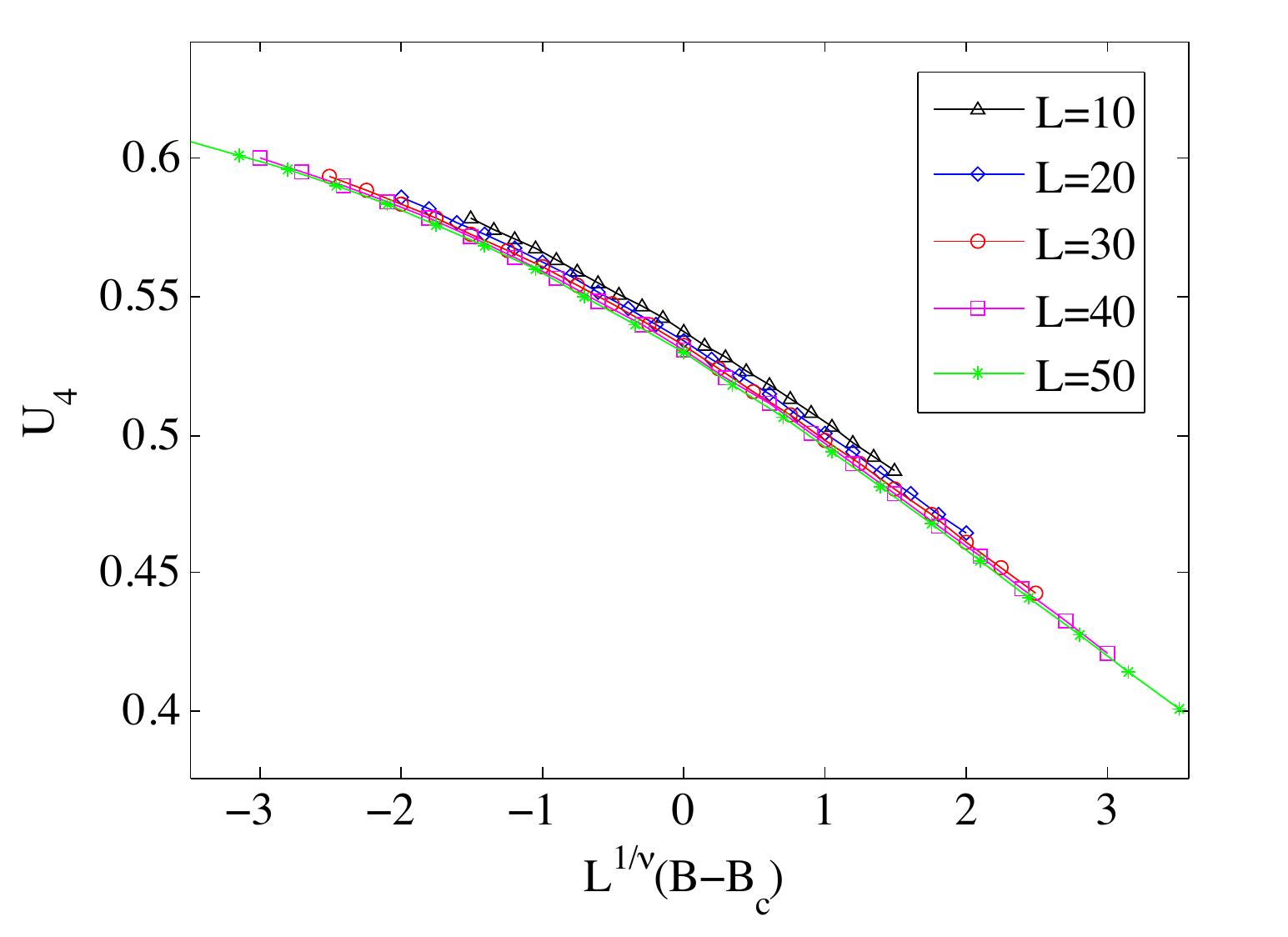}
\caption{\label{fig:CrystalCollapse} (Color online) The Binder cumulants for the spin-1 Ising model with crystal field, plotted for a variety of system sizes as a function of $L^{1/\nu}(B-B_c )$ for the known values $\nu = 1$ and $B_c = 2$. As expected, for these values the curves are seen to collapse to a functional form essentially independent of the length scale. } 
\end{figure}

Direct examination of the second cumulant in the infinite system is also still a viable method for estimating the transition point. In this case, the location of the maximum gives $B_c = 1.996(1)$ (see Fig.~\ref{fig:CrystalM2x}). As before, it also remains possible to study the critical exponent $\nu$ and critical field value simultaneously by seeking to collapse the data to its universal behavior as a function of $L^{1/\nu}(B-B_c)$ (Fig.~\ref{fig:CrystalCollapse}).

\subsection{Spin-1/2 Ising model on a 2D lattice }
Finally, as a proof-of-principle, we briefly demonstrate the application of these methods to a two-dimensional system: the spin-1/2 Ising model on a 2D square lattice. This system is described by the Hamiltonian

\begin{equation}
H = -\sum_{j, k} \sigma_{j,k}^x(\sigma_{j+1, k}^x + \sigma^x_{j, k+1}) + B S_{j,k}^z,
\end{equation}
where the subscripts are understood to terminate at the boundary of the system.
As discussed above, the study of two dimensional systems with tensor network states is considerably more involved than the study of one-dimensional systems. More elaborate methods must be undertaken to numerically approximate the ground states, and elaborate approximation schemes must be used in order to calculate expectation values, which are otherwise prohibitively costly in time and memory. A detailed and high-precision study of the critical point of this model is therefore beyond the scope of this paper (see instead \cite{2DIsingCirac}). Nevertheless, we include the following rough estimation in order to demonstrate how easily the moment-generating function method can be generalized to higher dimensional states, as well as to underscore the utility of Binder cumulant techniques even for data calculated relatively cheaply.

To this end, we generate approximate ground states for the model, using a simple method of local updates (a 2-D generalization of TEBD) and the smallest nontrivial bond dimension, $\chi = 2$. To check the behavior, we consider the order parameter

\begin{equation}
M = \sum_{j,k} \sigma^x_{j,k}.
\end{equation}

\begin{figure}[ht]
\includegraphics[width = 90mm]{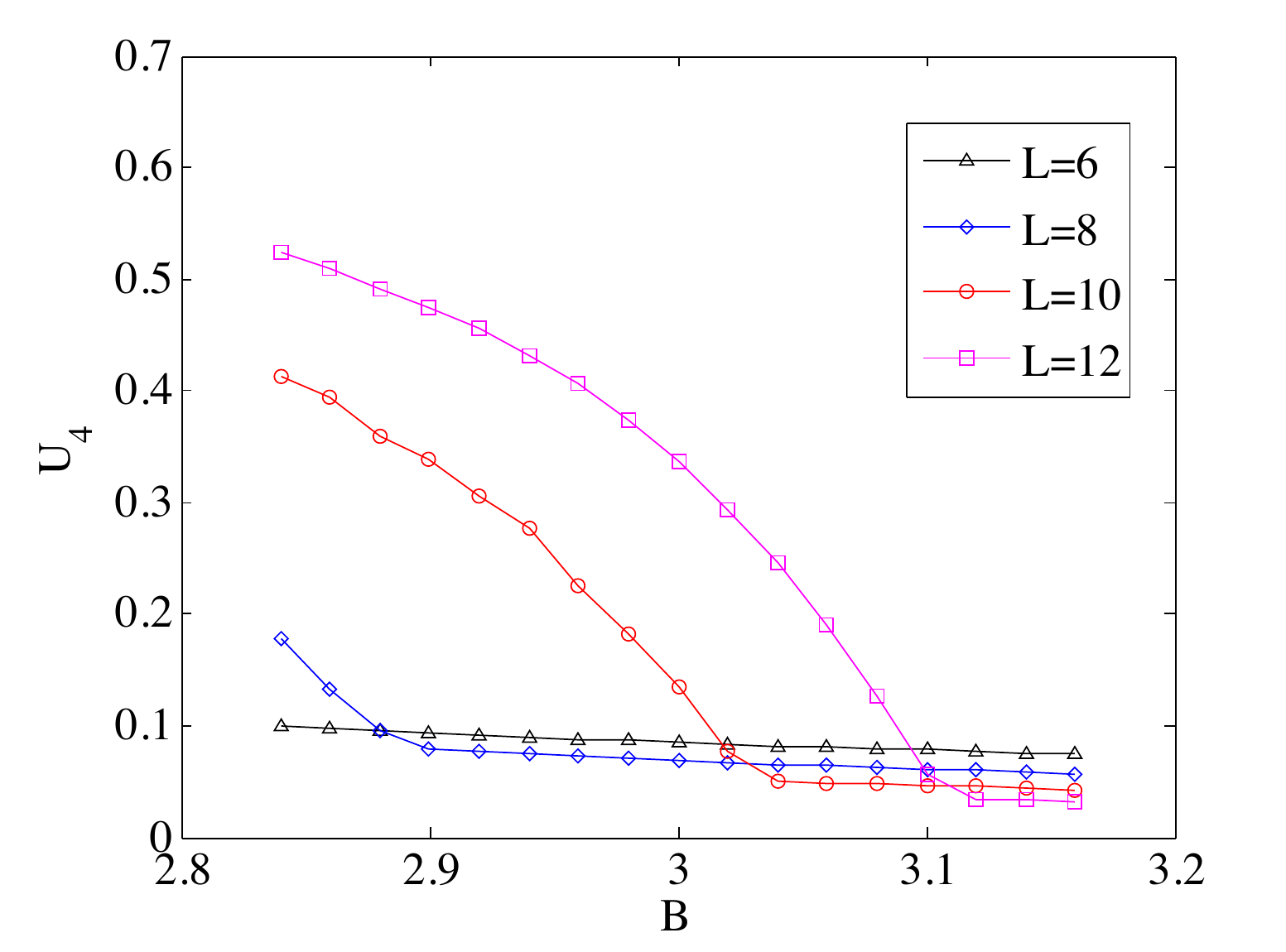}
\caption{\label{fig:Binder2DIsing} (Color online) A rough Binder cumulant study of the spin-1/2 transverse Ising model on a square lattice, using a local-update numerical algorithm with bond dimension $\chi = 2$. As in the one-dimensional case, the cumulants are computed for different system sizes across a range of values for the transverse field $B$. The largest crossing point, between $L=10$ and $L=12$, is at $B = 3.11(1)$. Extrapolating to the infinite limit gives $B_c = 3.3(2)$, though this cannot be done with high reliability on such a limited dataset (see text). }
\end{figure}

Then, as in the one-dimensional case, we compute the Binder cumulant of the order parameter across a range of applied fields, for systems of size $L \times L$  up to $L = 12$, and observe the crossings (Fig.~\ref{fig:Binder2DIsing}). The largest crossing we are able to compute, between $L = 10$ and $L = 12$, occurs at $B = 3.11(1)$, which is already reasonably accurate compared to the accepted value of $B_c = 3.044$, as calculated by quantum Monte Carlo \cite{2DIsingQMC}. A BST Extrapolation of the data gives $B_c = 3.3(2)$, a crude estimation but with relatively large error bars, owing largely to the fact that only three crossing values have been used in the extrapolation. We note also that, unlike the case of one-dimensional systems, the Binder crossings for two-dimensional systems are not necessarily converging monotonically and hence may not necessarily admit an easy extrapolation \cite{BinderSlides}. Instead, greater precision could likely be obtained through the use of more sophisticated two-dimensional methods (or additional computational resources) to study the crossings for slightly larger system sizes.

\section{Summary}

In this paper we have presented a method for efficiently calculating the higher order moments and cumulants of general operators on systems represented by tensor network states. For finite systems, this capability has a variety of applications in the search for phase transitions in quantum systems. Chief among these is the calculation of the celebrated ``Binder cumulant," which provides a powerful tool for not only detecting phase transitions, but determining their location to a high degree of accuracy using only relatively small finite systems to probe the infinite limit. The finite size scaling of the Binder cumulant also provides an estimate the critical exponent of the correlation length. Although the second cumulants of Hamiltonians have been considered in the context of matrix product states, to our knowledge, critical point detection techniques based on the Binder cumulant (or cumulants in general) have not generally been put to use in studies based on tensor networks, despite being widely applied to classical systems and quantum Monte Carlo studies. It is our hope that the methods presented in this paper will allow them to be embraced by the tensor network community as well.

In the case of infinite systems, we also present a method to calculate the per-site limits of the cumulants efficiently as well. The higher cumulants of an order parameter often show sharp behavior at the critical points, which in many cases allows for easier detect than the changes in the order parameter itself. In particular, we show how singularities in the second cumulant can produce a relatively precise (computationally cheap) estimation of the location of the transition. All the techniques (finite and infinite) are demonstrated in the context of the transverse Ising model, the spin-1 transverse Ising model, the Ising model in a crystal field, and could easily be applied to other models. We also demonstrate a useful application of the second cumulant of the energy. This quantity, which we calculate for both finite and infinite systems, provides a useful sufficient condition to determine when a numerical ground-state estimation algorithm has converged. As we demonstrate in the context of the Ising model, it can identify convergence up to a very high level of precision.

Finally, we present a proof-of-principle demonstration of the methods as applied to the transverse Ising model on a square lattice. Our result demonstrates that the methods for computing moments and cumulants are easily generalized to states in two dimensions or higher. Precise calculation of the moments and cumulants of a such a system may be more difficult, since state preparation and the process of computing expectation values are themselves much more complicated in higher dimensions. However, the central idea of our method on its own remains just as straightforward as in one dimension.

During the preparation of this work, it was brought to our attention that a technique based on matrix product operators (MPOs) has also been suggested as an alternative method for evaluating cumulants and moments, particularly in the context of the second cumulant of the energy and its use as a convergence check. We believe both methods have complimentary strengths and weaknesses, depending on the context in which they are applied.

\section{Acknowledgements}
The authors are grateful to Ching-Yu Huang for many useful discussions, to Frank Pollmann, who drew our attention to the matrix product operator method, and to Volckmar Nebendahl, who demonstrated to us the possibility of an MPO method which scales linearly. This work is supported by the National Science Foundation under grants No. PHY-1314748 and No. PHY-1333903.

\bibliography{BinderProjectDraft7}{}
\bibliographystyle{unsrt}

\appendix
\section{Extrapolation with the BST Algorithm}

We now briefly overview the Bulirsch-Stoer extrapolation scheme, commonly referred to as the ``BST" Algorithm (the meaning of the ``T" in this acronym has evidently been lost to time). This method was introduced in \cite{BSTOriginal} in the context of differential equations, but has been widely adopted as an extrapolation scheme whenever one seeks to project a sequence of data with unknown functional form to its infinite limit. In particular it has become a useful tool for finite-size scaling techniques, and was studied extensively in the context of lattice models in \cite{BSTLattice}. 

The BST Algorithm assumes that we are attempting to extrapolate to a limiting value for an infinite system, which is subject to power-law corrections when approximated by a finite-size. For example suppose we have a sequence of critical field values which are approximants to the true value of the critical field in the infinite system:  $\{B(L_1), B(L_2)...B(L_N)\}$, which approach $B_\infty \equiv B(L \to \infty)$. The BST Algorithm applies when, for each estimate $B(L)$, $B_\infty - B(L) = P(L)$ for some fixed (but unknown) polynomial $P$. This pattern of power-law corrections has generally been found to be true in the case of Binder Cumulants \cite{PowerLaw1, PowerLaw2}.

The technique works by taking the initial sequence $\{ \alpha^{(0)}_1, \alpha^{(0)}_2, ... \alpha^{(0)}_N\}$ and using it to construct a new sequence,  $\{ \alpha^{(1)}_1, \alpha^{(1)}_2, ... \alpha^{(1)}_{k-1}\}$ whose convergence towards the infinite limit has been accelerated, so that $\alpha^{(1)}_{N-1}$ is in fact a better estimate than $\alpha^{(0)}_N$. For clarity, note that we are using parenthetical superscripts to label the sequence, and subscripts to enumerate the terms within a sequence.

 The terms in this new sequence are defined as follows

\begin{equation}\label{eq:BST}
\alpha^{(j+1)}_k \equiv \alpha^{(j)}_{k+1} + \frac{\alpha^{(j)}_{k+1} - \alpha^{(j)}_k}{\left( \frac{L_{k+1}}{L_k} \right)^ \omega \left( 1- \frac{\alpha^{(j)}_{k+1} - \alpha^{(j)}_k} {\alpha^{(j)}_{k+1} - \alpha^{(j-1)}_{k+1}} \right) -1}.
\end{equation}

Note that, since the denominator of equation \ref{eq:BST} makes reference to the sequence $\alpha^{(j-1)}$, it is necessary to define the sequence $\{\alpha^{(-1)}\}$ to handle the initial step of the algorithm in which $j = 0$. To that end, one simply takes $\alpha^{(-1)}_k = 0$ for all $k$.

This procedure can then be repeated, taking the sequences $ \alpha^{(0)}$ and $\alpha^{(1)}$ as the inputs to generate $\alpha^{(2)}$, and so on. This iteration can be done at most $N-1$ times, at which point the resulting sequence $\alpha^{(N-1)}$ contains only one term. This term is the BST algorithm's best estimation of the infinite limit of the original sequence.

The parameter ``$\omega$", which appears as the exponent on the length scales, is a free parameter in the algorithm. The value of $\omega$ which gives the best convergence will depend on the form of the power law corrections in the original sequence, which is generally unknown. Hence, in practice, a range of parameters must be considered, selecting the one which best optimizes the convergence. To this end, note that a rough estimate of the ``precision" of the sequences can be made by

\begin{equation}
\Delta^{(j+1)}_k = 2 | \alpha^{(j)}_k - \alpha^{(j)}_{k+1} |.
\end{equation} 

This value should be decreasing with each iteration if the procedure is correctly accelerating the convergence of each new sequence. The final value of this estimator, $\Delta_{final} = \Delta^{(N-1)}_1$ gives a convenient way to fix the free parameter $\omega$: we simply repeat the algorithm while varying $\omega$, and choose the one which minimizes $\Delta_{final}$. In practice, it has often been observed \cite{BST2} that the dependence of the estimations on $\omega$ is rather weak, with large ranges of values giving comparable results. In other words, it is often more important simply to avoid a ``bad" value of $\omega$ than to try to find its absolute ``best" value. In our work, we have used as our procedure a sweep over the range $\omega \in (0, 2]$, testing with steps of size $0.1$. We also require that our extrapolation be "stable" under small variations in $\omega$.  

We note that the value of $\Delta_{final}$ cannot be used as a complete measure of the error in a final estimation. It is a measure of the internal consistency and the precision of the acceleration in the BST algorithm, but cannot contain any information about whether the algorithm has captured the "true" functional form of the finite-size corrections. Additionally, it does not reflect the propagation of errors on the data which are being extrapolated. A very small and stable value of $\Delta_{final}$ indicates that the algorithm is extrapolating the data to the best of it's capability given its assumptions and the finite number of input points. It does not necessarily indicate that the result is extremely precise.

In this paper, to estimate the error in a BST extrapolation, we start with the uncertainties of the input points, and essentially determine the propagated uncertainty empirically. In our case, the input points are the crossings of Binder cumulant curves. The crossings are computed by linearly extrapolating between data points, so we generously assume an uncertainty of one half the step size between points. The error is then estimated by considering a ``worst-case scenario" in which the first few crossing points are perturbed downward by this amount, and the later points perturbed upwards. We run these perturbed points through the BST algorithm and observe the effect on the resulting extrapolation. The size of this effect is taken to be a rough upper bound on the total uncertainty.

\nocite{*}

\end{document}